\documentclass[aip,reprint, floatfix]{revtex4-1}

\usepackage{graphicx}

\usepackage[english]{babel}
\usepackage{amsmath}
\usepackage{float}
\usepackage{subcaption}
\usepackage[colorlinks=true, allcolors=blue]{hyperref}
\usepackage{physics}
\usepackage{sidecap}
\usepackage{siunitx}
\usepackage{diagbox}
\usepackage{nameref}
\usepackage{placeins}
\usepackage[normalem]{ulem}

\usepackage{overpic}
\usepackage{calc}
\usepackage{tikz}

\usepackage{amssymb}
\usepackage{latexsym}

\newlength{\height}
\newsavebox{\tbox}

\begin{document}


\title{Neural Networks for the Analysis of Traced Particles in Kinetic Plasma Simulations} 



\author{G. Torralba Paz}
\email[E-mail:]{gtorralba@ifj.edu.pl}
\affiliation{Institute of Nuclear Physics Polish Academy of Sciences, PL-31342 Krakow, Poland}

\author{A. Bohdan}
\affiliation{Max-Planck-Institut für Plasmaphysik, Boltzmannstr. 2, DE-85748 Garching, Germany}
\affiliation{Excellence Cluster ORIGINS, Boltzmannstr. 2, DE-85748 Garching, Germany}

\author{J. Niemiec}
\affiliation{Institute of Nuclear Physics Polish Academy of Sciences, PL-31342 Krakow, Poland}

\date{\today}

\begin{abstract}
    Cosmic-ray acceleration processes in astrophysical plasmas are often investigated with fully-kinetic or hybrid kinetic numerical simulations, which enable us to describe a detailed microphysics of particle energization mechanisms. Tracing of individual particles in such simulations is especially useful in this regard.
    However, visually inspecting particle trajectories introduces a significant amount of bias and uncertainty, making it challenging to pinpoint specific acceleration mechanisms. 
    Here, we present a novel approach utilising neural networks to assist in the analysis of individual particle data. We demonstrate the effectiveness of this approach using the dataset from our recent particle-in-cell (PIC) simulations of non-relativistic perpendicular shocks that consists of 252,000 
    electrons, each characterised by their position, momentum and electromagnetic field at particle's position, recorded in a time series of 1200 time steps. These electrons cross a region affected by the electrostatic Buneman instability, and a small percentage of them attain high energies. We perform classification, regression, and anomaly detection algorithms on the dataset by using a convolutional neural network, a multi-layer perceptron, and an autoencoder. Despite the noisy and imbalanced dataset,  
    all methods demonstrate the capability to differentiate between thermal and accelerated electrons with remarkable accuracy.
    The proposed methodology may considerably simplify particle classification in large-scale PIC and hybrid simulations.
\end{abstract}


\maketitle 

\section{Introduction}
Cosmic rays are charged non-thermal particles accelerated in astrophysical plasmas
of various galactic and extragalactic sources, in the latter reaching energies as high as $10^{21}$~eV. Acceleration processes in these sources shape particle distributions that are responsible for the observed radiation emission spectra and the flux of cosmic rays detected at Earth.

Particle acceleration processes in astrophysical plasmas can be best investigated in detail using fully kinetic Particle-in-Cell (PIC) or hybrid kinetic numerical simulations~\citep{PIC}. Such simulations model the motion of billions of particles interacting with self-generated electromagnetic fields. They provide integrated particle distributions, offering valuable insights into particle heating and acceleration efficiencies. However, for a detailed understanding of the underlying mechanisms of particle energization, the key advantage of kinetic simulations is their ability to trace individual particle trajectories within the plasma, allowing one to pinpoint the exact acceleration processes affecting each particle.

Analyzing particle tracing data from kinetic simulations is typically done by visually inspecting numerous particle orbits. However, this manual approach introduces significant bias and uncertainty, which can lead to inconsistent or unreliable results. In this work, we propose a novel method for analyzing particle tracing data using machine learning, specifically Neural Networks (NN).

NNs offer the capability to swiftly and reliably postprocess thousands of particles, facilitating the study of the underlying acceleration mechanisms. Each process exhibits a unique behaviour in the variation of particle and field parameters -- a fingerprint that can be identified by NNs. NNs have previously been applied in high-energy astrophysics for identification of neutron star mergers \citep{mergers}, in image processing in Cherenkov telescopes \citep{gr_detector}, and parametrization of neutrino detection in IceCube \citep{neutrinos}.

In our work, we propose a novel application of NNs. We develop and test the particle tracing data analysis methods based on the dataset taken from our recent PIC simulations, and employ three different algorithms (classification, regression and anomaly detection) based on two NN types: Multi-layer Perceptron (MLP) and Convolutional Neural Networks (CNN). To evaluate the applicability of NNs as an analysis tool, we consider the Buneman instability\citep{Buneman} as a test case. This widely-studied instability is known for its role in accelerating electrons\citep{ShimadaHoshino, Hoshino2002}.

The paper is organized as follows. Section 2 presents the traced particle dataset obtained from our kinetic PIC simulations and describes a specific acceleration process that these particles underwent. Section 3 details NNs and algorithms used for analyzing the dataset. Section 4 presents the results, while Section 5 summarizes these results and explores the potential for applying the developed techniques to future analyses.

\section{Description of the dataset}
For our study we select a case of nonrelativistic perpendicular shocks of young supernova remnants, which have been recently explored with numerous fully-kinetic and hybrid-kinetic simulations \cite[e.g.,][]{Matsumoto_2012, Bohdan2017, runE2, Bohdan_Buneman}. These shocks are characterised by high Alfvénic and sonic Mach numbers and propagate in weakly magnetized low-temperature interstellar medium. The physics of such shocks
involves a small population of incoming ions that are reflected off the shock front back upstream, where they interact with the upstream thermal plasma, generating a non-zero current that drives the pre-shock plasma unstable. One of the instabilities that occur in perpendicular and oblique high Mach number shocks is the Buneman instability \cite[e.g.,][]{buneman_amano,Matsumoto_2012, Bohdan_Buneman}, which is a type of kinetic instability that is excited in the shock foot through the relative drift between ions reflected by the shock potential and upstream electrons. The instability produces electrostatic waves. Some upstream electrons that interact with these waves are accelerated to high energies via the so-called shock-surfing acceleration (SSA) process \citep{Hoshino2002}.
 
\begin{figure*}[ht!]
    \centering
    \includegraphics[width=0.99\linewidth]{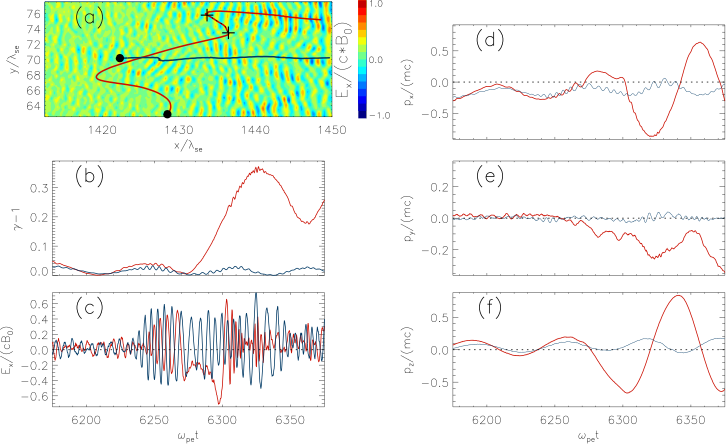}
    \caption{Interaction of two distinct sample electrons with the Buneman waves in the shock foot. The electron for which the data are shown in red is accelerated via SSA, whereas the dark blue electron passes the Buneman instability region effectively unaffected. Panel (a): the map of the $E_x$ electric field component at time $t\omega_{pe}=6375$. 
    Overlaid are the positions of the electrons at the same moment (black dots), and their trajectories over the past $200\omega_{pe}^{-1}$. The two plus signs indicate the period of trapping of the red-lined electron. Panel (b): evolution of the kinetic energy of electrons, $\gamma-1=E_k/m_ec^2$. Panel (c): normalized amplitude of $E_x$ electric field at electron positions in the simulation frame. Panels (d-f): evolution of electron momenta components.}
    \label{fig:bun_traj}
\end{figure*}

The dataset utilised in this study was obtained from two-dimensional (2D) PIC simulation run E2 as described in \onlinecite{runE2}. 
This run is representative in terms of the Buneman wave structure and amplitude, and the SSA efficiency at a typical high Mach number shock, whose parameters satisfy conditions
for efficient electron pre-acceleration. Run E2 simulates a shock with an Alfvénic Mach number of $M_A=44.9$ and a sonic Mach number of $M_s=69$, propagating through an electron-ion plasma with an electron plasma beta (the ratio of the electron plasma pressure to the magnetic pressure) $\beta_e=0.5$ over time-span of $t=10500\,\omega_{pe}^{-1}= 420,000 \delta t$, where $\omega_{pe}$ denotes the electron plasma frequency and $\delta t$ is the simulation time step. The simulated plasma is contained on a 2D Cartesian grid in the $xy$ plane, with the shock propagating along the $x$-direction. The simulation tracks two spatial coordinates, along with all three components of particle velocities and electromagnetic fields.
In the adopted geometry, a homogeneous magnetic field lies in the $y$-direction, $\mathbf{B}_0=B_{y0}\hat{y}$, perpendicular to the shock normal. In this configuration, the wavevectors of the Buneman waves align approximately along the $x$-axis. Hence, these electrostatic waves are mainly seen in the $E_x$ component of the electric field, as illustrated in Figure~\ref{fig:bun_traj}a, and the $E_y$ and $E_z$ field components remain constant throughout the region. 

We collected particle tracing data for run E2 from $N=252,000$ electrons, recording their positions ($x$ and $y$), momentum components ($p_x$, $p_y$, and $p_z$) as well as the magnetic field, $\mathbf{B}=(B_x,B_y,B_z)$, and the electric field, $\mathbf{E}=(E_x,E_y,E_z)$, at their respective positions. 
Selected electrons were traced for a total duration of $t=4500\,\omega_{pe}^{-1}$. From this tracing data, a segment of $300\,\omega_{pe}^{-1}$ was extracted, corresponding to the period during which the electrons moved through the Buneman instability region and interacted with electrostatic waves. For the analysis, we use this time series of $t=300\,\omega_{pe}^{-1}$ equivalent to $=12,000\,\delta t$ with a sampling interval of $0.25\,\omega_{pe}^{-1}=10\,\delta t$.

Figure \ref{fig:bun_traj}a shows two typical and distinct trajectories of electrons traversing the Buneman instability region, overlaid on a map of the $E_x$ electric field. These electrons were traced over the time interval from $t=6075\,\omega_{pe}^{-1}$ to $t=6375\,\omega_{pe}^{-1}$. In
Figure~\ref{fig:bun_traj}, we present the final $200\,\omega_{pe}^{-1}$ of their traced
trajectories.
The electron whose trajectory is shown with the blue solid line
experiences slight heating due to the electrostatic waves and primarily gyrates around the average magnetic field.
The other electron (trajectory depicted with the red solid line) undergoes SSA, which results in a nearly tenfold increase in its kinetic energy
(Figure~\ref{fig:bun_traj}b).
Energy gain for this electron occurs during the time interval between $t\omega_{pe}\approx 6275$ and $6300$, indicated by two plus signs in Figure~\ref{fig:bun_traj}a, when the particle is trapped by a wave. This trapping period is characterized by a nearly constant (non-oscillatory) negative $E_x$ electric field at the electron's position (Fig.~\ref{fig:bun_traj}c), which is the unique "fingerprint" of the SSA process. During this stage the electron moves together with the elecrostatic wave in the $x$-direction with $p_x/m_ec\approx 0.2$ (Fig.~\ref{fig:bun_traj}d), its $p_y$ momentum remains small ($p_y/m_ec\approx 0.1$, Fig.~\ref{fig:bun_traj}e), and the electron is accelerated in the $z$-direction due to the motional electric field in the frame of the electrostatic wave, causing its $p_z$ momentum to increase to $p_z/m_ec\approx 0.6$ (Fig.~\ref{fig:bun_traj}f). At $t\omega_{pe}\approx 6300$ the electron resumes gyration in the magnetic field.
It is important to note that only a small fraction of particles undergo SSA and are accelerated to high energies, while the majority of electrons experience minor heating (see below).

Our analysis of particle trajectories with NN methods in search of electrons accelerated via SSA focuses on the characteristic features in particle momenta and $E_x$ electric field described above. 

\begin{figure}[ht!]
    \centering
    \includegraphics[width=\linewidth]{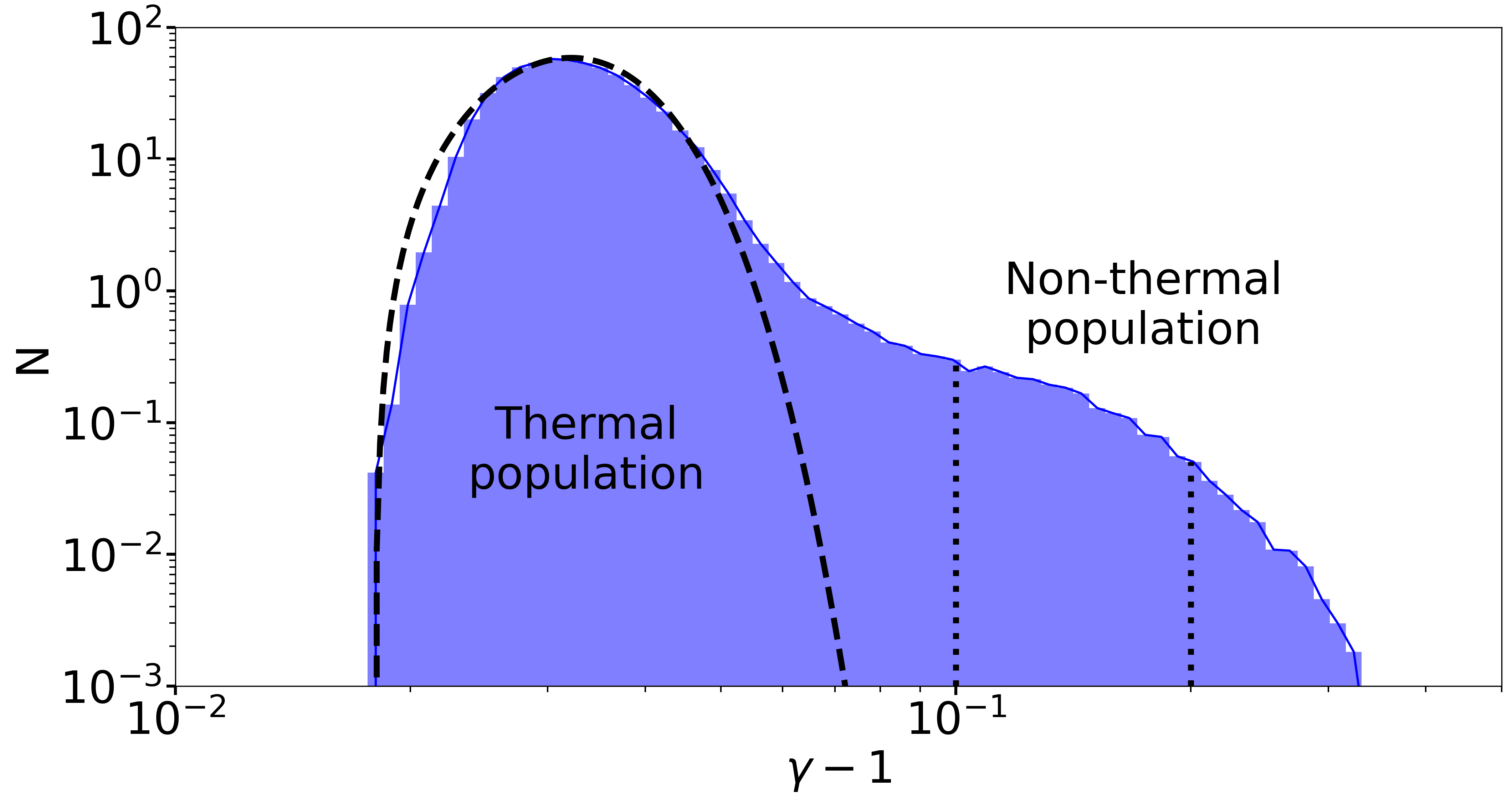}
    \caption{Histogram of the maximum kinetic energy of the particles in our dataset. The bulk of the particles forms a thermal population, approximately marked with the dashed line, 
    whereas around 2\% of particles have been energised and belongs to the non-thermal population (right part of the histogram). The vertical dotted lines split the histogram in the thermal, supra-thermal, and high-energy populations, as defined in Table~\ref{tab:classes}.}
    \label{fig:hist}
\end{figure}

We assign a label $y_i$ for each particle that corresponds to the maximum kinetic energy achieved along its path in the Buneman instability region:
\begin{equation}
    y_i=\max{(\gamma_i-1)},\quad \gamma=\sqrt{1+\left(|\vb{p}|/(m_ec)\right)^2},
    \label{eq:max_en}
\end{equation} 
where $\gamma_i$ is the Lorentz factor of the $i$-th particle, $|\vb{p}|$ is the momentum magnitude, $m_e$ the electron mass, and $c$ the speed of light. 
The distribution of the maximum kinetic energy within our dataset is presented in Figure~\ref{fig:hist}. 
One can see that the bulk of the electrons form a population with an average Lorentz factor of $\gamma-1\simeq 0.03$. These are the incoming upstream electrons thermalized by the Buneman instability waves. In the following, we refer to these electrons as to the \emph{thermal} population.
Approximately 2\% of electrons reach maximum kinetic energies exceeding $\gamma-1=0.1$. This notable imbalance in the dataset will have implications for the analysis and will be addressed further in this work. Our primary objective is to identify and characterise these high-energy particles. 
These electrons, having been pre-accelerated by SSA, may interact again with the shock front and undergo further acceleration by other mechanisms, which is not viable for thermal particles due to their insufficient kinetic energy. 

\section{Algorithms and Methods}
\subsection{Neural Networks}
In this section, we present two fundamental neural networks used in the analysis of our dataset: the convolutional neural network (CNN) and the multi-layer perceptron (MLP). 
We have selected these two neural networks for their specific capabilities. CNNs are known for their usefulness in handling 2D and 1D arrays as input data and are widely used for tasks such as image classification and object detection \citep{time_series}. CNNs employ convolutional layers to extract spatial and temporal patterns within datasets, taking advantage of the locality inherent in the convolution operation. These convolution layers are fundamental in extracting localised time patterns that influence the time series analysed in our study. Specifically, these patterns correspond to acceleration phenomena caused by electrons interacting with Buneman electrostatic waves. The MLP, as the most elementary form of NN, is an ideal choice for directly comparing with the CNN, providing insights into their respective performance capabilities.

\subsubsection{Convolutional Neural Networks} \label{sec:cnn}
In a CNN, the core components are convolutional layers, each comprising of multiple filters or units. These filters have a convolutional kernel with a given window size (e.g., 1D for time series data, 2D for images) that are filled with weights. These windows slide through the input data performing convolutional operations.
The dimension of the output as a result of a convolutional layer is determined by the number of filters within that layer. Typically, the dimension of the output space is lower than that of the input space.

\begin{figure*}
    \centering
    \includegraphics[width=0.9\linewidth]{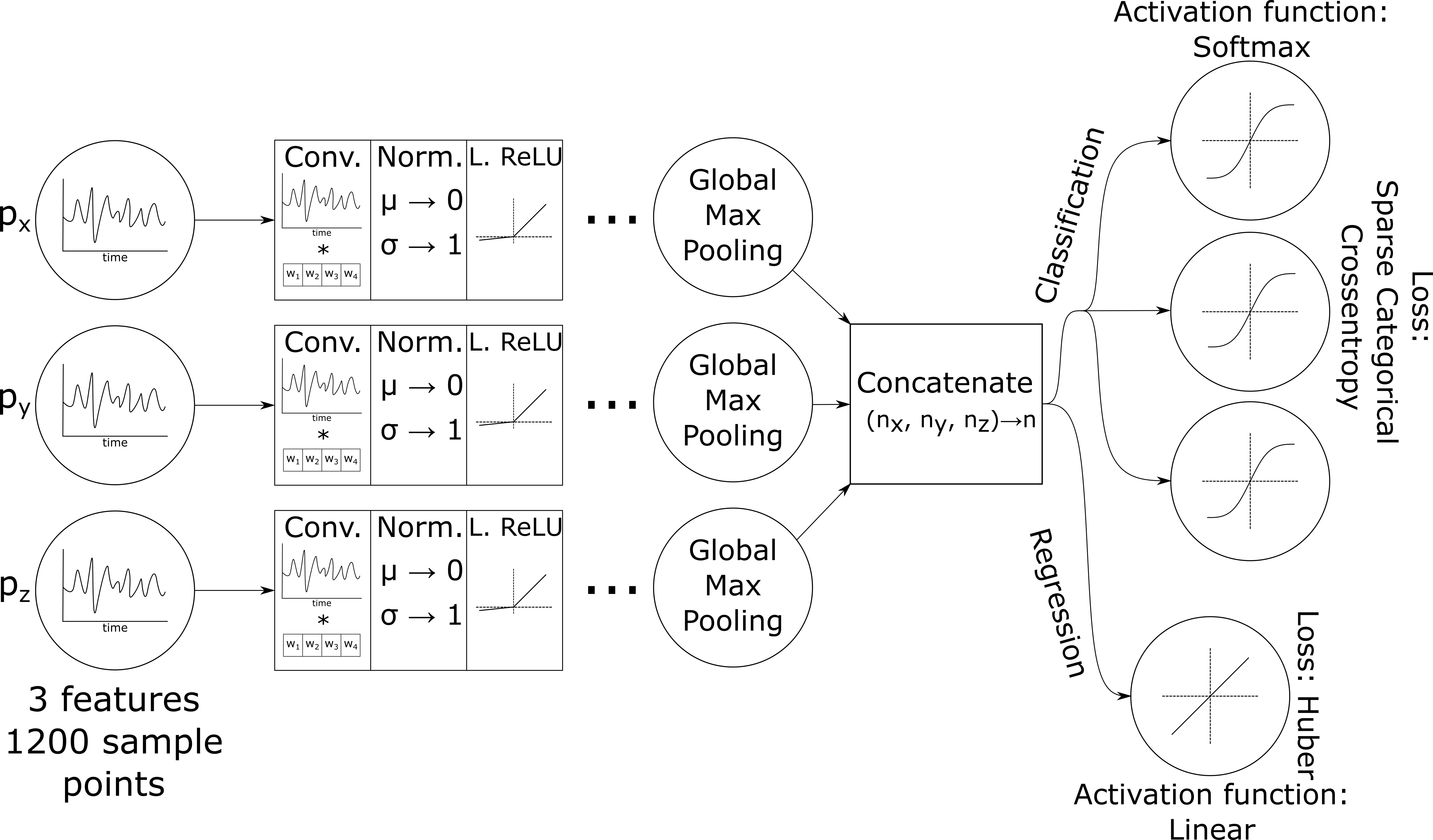}
    \caption{A simplified schematic version of a CNN,
    using the regression algorithm as an example (see Section \ref{sec:reg}). The input data (time series of $p_x$, $p_y$ and $p_z$) passes through several sets of layers (denoted by the rectangular window), each containing a convolutional layer
    (with the convolutional window weights denoted by $w_i$), a standardising layer that sets the average value to 0 ($\mu\rightarrow0$) and the standard deviation to 1 ($\sigma\rightarrow1$), and a non-linear activation function, in this case, a Leaky ReLU. After the information is processed through all these sets of layers denoted by an ellipsis, the Global Max Pooling is applied to the output and and the results are concatenated. Finally, the result passes through the output layer, which in this case is a single neuron with a linear activation function.}
    \label{fig:cnn}
\end{figure*}

The CNN architectures we use in this work consist of either three branches, with each branch corresponding to one of the three momentum components, or a single branch, in the case the input variable is the $E_x$ electric field component.
Figure~\ref{fig:cnn} shows a simplified representation of the three-branch CNN that has particle momentum components provided as inputs. Each branch of our CNN is composed of multiple sets of layers, each containing three distinct layers in a sequence, represented as a rectangular window in the figure. These include a convolutional layer that applies convolutional operations to the input data using a specific kernel, a standardisation layer to optimize data processing within the CNN, and a Leaky Rectified Linear Unit (Leaky ReLU) activation function \citep{relu}, which imparts the NN with its characteristic non-linear behavior to enable it to learn complex relationships in the data. Following these layers, the output passes through a Global Max Pooling layer, which retains the maximum value along the time dimension for each filter. 
An Average Pooling layer could have been used as well, but the Global Max Pooling layer performs better for the specific problem addressed in this work. After the Global Max Pooling step, the results from the three branches are concatenated and flattened. At this stage in the CNN, the specific design of the final layer depends on the particular algorithm under consideration. For instance, in regression tasks, a single neuron is used, whereas in classification tasks, the number of neurons matches the number of classes in the dataset.

\subsubsection{Multi-Layer Perceptrons}
Multi-Layer Perceptrons (MLPs) are one of the most simple NNs and more general-purpose than CNNs. They consist of a fully connected network of layers, with the size determined by the number of units. Each of these units is characterised by a weight and an activation function which, instead of conducting a convolution, performs a weighted sum over the input values. As in the case of CNNs, we use Leaky ReLU as the activation function for the MLPs.

\begin{figure*}
    \centering
    \includegraphics[width=0.8\linewidth]{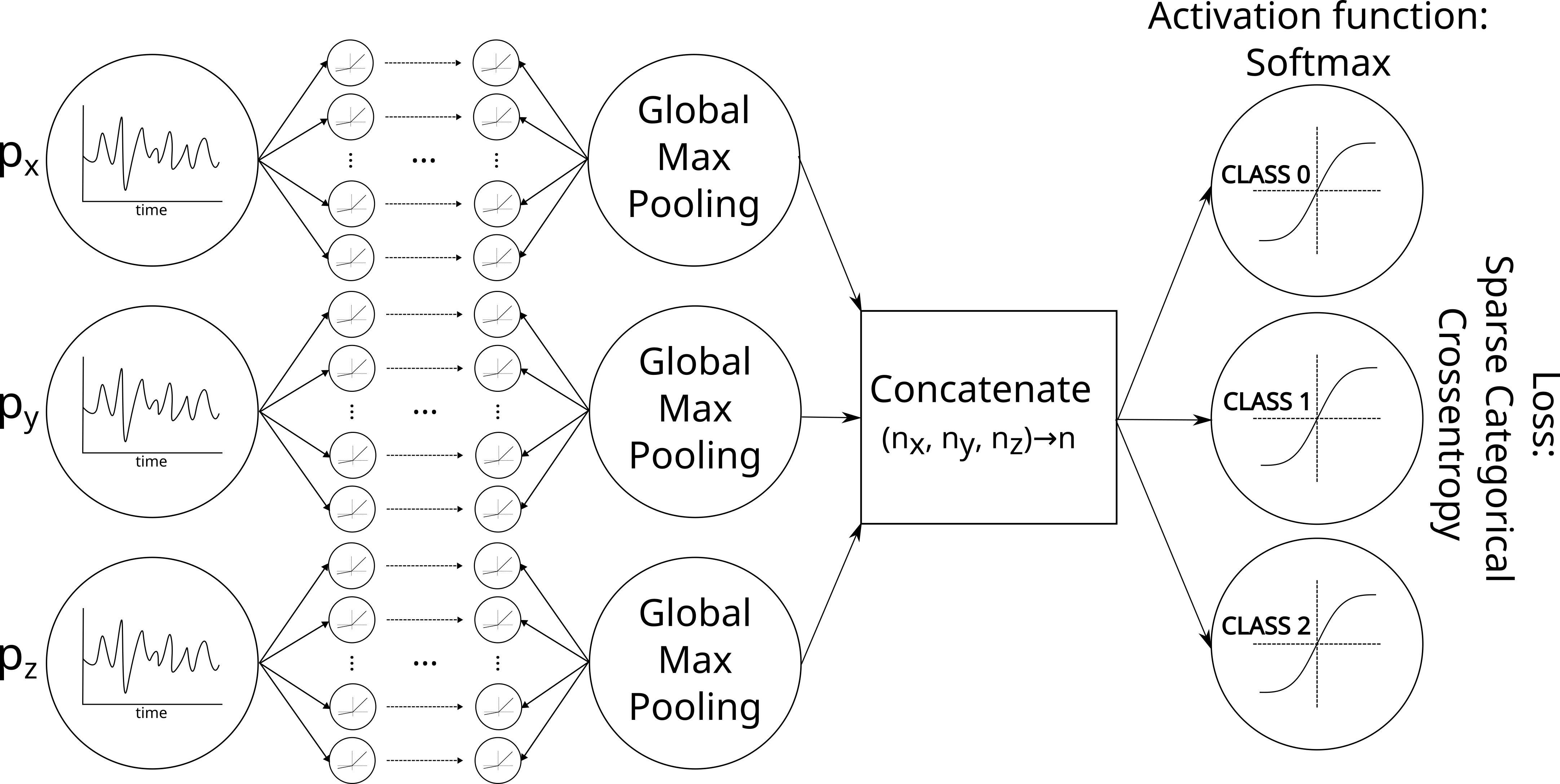}
    \caption{Simplified version of a MLP with the classification algorithm (see Section \ref{sec:cls}). Each input feature passes through a series of fully-connected layers, denoted by small circles. Subsequently, the data goes through a Global Max Pooling layer. Subsequently, the data from the three features is concatenated and flattened. Finally, the data goes through the output layer, which in this case consists of three neurons, each with a softmax activation function that performs classification.}
    \label{fig:mlp}
\end{figure*}

Figure \ref{fig:mlp} shows a visual representation of a MLP used in this work.
Like in our CNNs, the MLPs consist of individual branches that correspond to distinct input features -- in the application considered here they are either three momentum components or $E_x$ electric field. Each branch is composed of a fully connected network. Ultimately, the results from all branches are concatenated, flattened, and passed to the output layer, following the same approach as in our CNNs. 

\subsection{Algorithms}
\subsubsection{Classification} \label{sec:cls}
Classification is a supervised machine learning task, whose goal is to categorize input data into predefined discrete classes or labels. There exist two distinct categories: mutually exclusive classes, in which each input data is exclusively assigned to a single class, and mutually inclusive classes, in which the data can belong to multiple classes simultaneously. In our specific application, we use mutually exclusive classes only. 

The output of the classification algorithm consists of $k$ neurons, where $k$ is the number of classes or labels defined for the dataset. Each neuron has a softmax activation function applied to its output \citep{deep_learning}, which transforms the input data in the form of a real vector $\vb{x}$ with $k$ elements into another real $k$-element vector $\vb{\sigma}(\vb{x})$, defined as:
\begin{equation}
    \vb{\sigma}(\vb{x})_i = \frac{e^{\vb{x}_i}}{\sum_{j=1}^ke^{\vb{x}_j}},\quad i=1,...,k.
\end{equation}
The normalization ensures that the output values lie between 0 and 1 and that they sum up to 1. This normalization thus transforms the output of a NN into a probability distribution over multiple classes, which makes the softmax activation function useful in classification problems.

For our applications, we split our dataset into three distinct classes, each corresponding to specific intervals of the maximum kinetic energy, as delineated in Table \ref{tab:classes}. This categorization somewhat arbitrarily defines the \emph{thermal} particle population (Class 0), the \emph{supra-thermal} population (Class 1), and the \emph{high-energy} particle population (Class 2; compare Figure~\ref{fig:hist}). An example of a NN with the classification algorithm with 3 classes and the softmax activation function is shown in Figure~\ref{fig:mlp}.
To assess the accuracy of the classification algorithm, we use the Sparse Categorical Cross-Entropy loss function \citep{Bishop}, a widely employed loss function that efficiently handles multi-class classification tasks with integer-encoded labels.

\begin{table}[ht]
    \caption{Maximum particle kinetic energy ranges for which we define classes in our dataset. Also shown are the numbers of particles in each class for our dataset. Note the imbalance in the number of particles per class.}
    \label{tab:classes}
    \centering
    \begin{tabular}{c c c}
        \hline\hline
        Maximum Kinetic Energy & Class & No. particles \\
        \hline
        $\gamma-1 \leq 0.1$       & 0     & 247722       \\
        $0.1 < \gamma-1 \leq 0.2$ & 1     & 3788         \\
        $\gamma-1 > 0.2$       & 2     & 490          \\
        \hline\hline
    \end{tabular}

\end{table}

\subsubsection{Regression} \label{sec:reg}
Regression is another supervised machine learning method in which a NN is employed to make predictions about a continuous-valued feature in the input data. This means the feature can assume any real numerical value. The output layer of such NN consists of a single neuron to which a linear activation function $y=x$ is applied, as schematically depicted in the output layer of Figure~\ref{fig:cnn}. The output value generated by this single neuron is the label prediction, which in our application is the maximum kinetic energy attained by a given particle. This prediction is compared to the original value, $y_i$ (see Equation~\ref{eq:max_en}) by using the Huber loss function \citep{huber}, which assesses the accuracy of our regression algorithm. This function combines the advantages of both mean absolute error (MAE) and mean squared error (MSE) loss functions, while demonstrating reduced sensitivity to outliers compared to MSE.

\subsubsection{Anomaly Detection}
\begin{figure*}
    \centering
\includegraphics[width=0.8\linewidth]{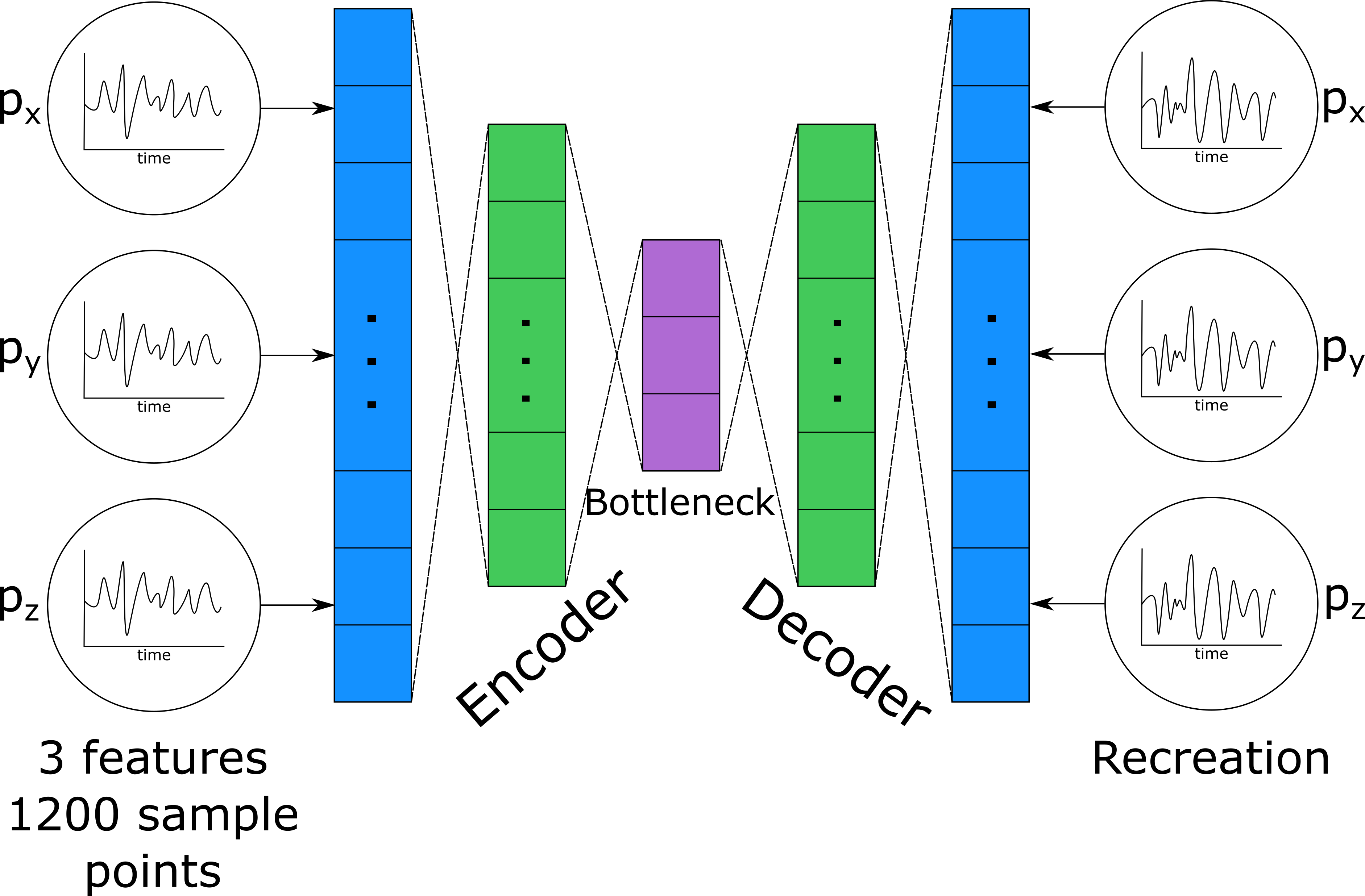}
    \caption{Simplified representation of an autoencoder which performs anomaly detection. The input data, represented here with the momentum components, is encoded and passes through several convolutional or fully-connected layers until it is compressed into a layer with a very small number of filters, denoted as the bottleneck. Finally, the compressed data is decoded and reconstructed to match the original input data.}
    \label{fig:ad}
\end{figure*}

Anomaly Detection (AD) is an unsupervised learning method designed to identify input values that differ from the norm. This algorithm is particularly well-suited for imbalanced datasets, in which infrequent values are categorised as outliers or anomalies. The autoencoder designed for AD and used in this work is illustrated in Figure~\ref{fig:ad}. This NN comprises a sequence of a series of convolutional layers or fully-connected layers that gradually increase in density, i.e., they contain fewer filters or units, effectively encoding the data.
At the midpoint of the autoencoder, a bottleneck is created, where the most crucial information is compressed into a very small number of filters or units. Subsequently, the autoencoder performs the reverse operation (e.g., deconvolution in the case of a CNN), effectively decoding the data by progressively increasing the amount of filters/units. Finally, the data passes through the output layer, which consists of a number of neurons equal to the number of input features. Each of these neurons contains the reconstruction of the input data. If the reconstruction closely resembles the original input data, it is categorised as non-anomalous, indicating a typical data in the dataset. Conversely, if the reconstruction significantly differs from the original data, the NN tags this data as anomalous, indicating data that is uncommon in the dataset. To distinguish between anomalies and non-anomalies, a threshold is computed by taking the highest (or close to the highest) loss value obtained during the training phase. During the testing phase, if for a given input data the loss is higher than the previously computed threshold, it is classified as an anomaly. We use the LogCosh loss function, which closely resembles the Huber loss function. 

\subsubsection{Hypertuning}
All NNs rely on a set of parameters known as hyperparameters, which include features like the number of layers, the number of neurons, the size of the convolutional window, etc. These hyperparameters play a crucial role in achieving optimal performance. Typically, they are manually tuned, a process that often involves trial and error to discover the best hyperparameters that result in the lowest validation loss.
The process of using algorithms to automatically search for the best hyperparameters is called hypertuning. In this work, the TPE Sampler algorithm \citep{tpesampler} from Optuna was employed for this purpose.
We note that hypertuning was applied to all of our NNs except for the ones used for regression with a CNN and anomaly detection.

\subsubsection{Data Weighting}
Because of the significant class imbalance in our dataset, the NN tends to make more accurate predictions for the time series that have more frequent labels,
e.g., for thermal particles. This is because the NN has been exposed to a larger number of 
these particles during training, in contrast to the energetic particles, which are much less frequent. This imbalance poses a challenge when performing classification or regression tasks, as it can lead to suboptimal results. To address this issue, we use initial weights during training. These initial weights are assigned to give greater importance to energetic particles
in the dataset.

\begin{itemize}
    \item Weighting in Classification: In the case of classification, 
    there are two parameters we can adjust before training begins to enhance the accuracy of the NN: the weights and biases of the output layer. A neuron computes its output as:
    \begin{equation}
        x_{out} = \mathcal{F}(\sum_i^{N}{(w_i\cdot x_i)}+b_i),
    \end{equation}
    where $\mathcal{F}$ is the activation function, $w_i$ and $b_i$ denote respectively the weights and biases for each input value $x_i$, and $N$ is the total number of input values with $N=3$ in our NN output layer. The weights can be calculated as the inverse of the frequency of each class, $f_i$, normalised to the total number of classes $n_{c}$:
    \begin{equation}
        w_i=n_{c}/f_i,\quad f_i=n_i/n,
    \end{equation}
    where $n_i$ is the number of particles in class $i$ and n is the total number of particles in the training dataset.
    
    In the case of biases, their calculation depends on the activation function used. For softmax activation, the biases are computed as:
    \begin{equation}
        f_i=\frac{e^{b_i}}{\sum_{j}^{n_c} e^{b_j}}.
    \end{equation}
    This computation results in a system of equations that typically requires numerical solution, unless $n_c=2$.

    \item Weighting in Regression: In the case of regression, there are no classes. Instead, we have a distribution of values as illustrated in Figure~\ref{fig:hist}. To address this, a weight is applied to each sample based on the frequency of the associated value. One approach of doing this involves smoothing the histogram using a smoothing kernel and then taking the inverse of the frequency depending on the value of the sample \citep{sample_weights}. Frequent values receive lower weights, while less frequent values receive higher weights. 
    
    Another method for handling imbalanced data involves transforming the labels to partially mitigate the imbalance. One such transformation is simply applying the logarithm to $\log{y_i}$, which yields a much flatter histogram.
\end{itemize}

\section{Results}
In Section~\ref{sec:results_momentum} we present the results obtained by using NNs and algorithms described above with input data in the form of the time series of the
three components of momentum, ($p_x, p_y, p_z$). Subsequently, in Section~\ref{sec:results_efield}, we conduct comparative investigations employing only time series of $E_x$ at particle positions as the input data.
These two sets of input data differ significantly from each other. Particle momenta are directly related to the kinetic energies of the particles, against which NNs perform various tasks. The momentum time series are also quite smooth (compare Fig.~\ref{fig:bun_traj}d-f). These characteristics facilitate the NN processing, and thus should result in more precise and sensitive predictions.
Conversely, the electric field data poses a challenge to NNs, since the $E_x$ time series obtained from simulations are characterized by a high level of noise (see Fig.~\ref{fig:bun_traj}c). Furthermore, the electric field along a particle orbit is not intrinsically linked to the kinetic energy of the particle. Consequently, we anticipate that NN perform better when provided with the particle momenta as input data, compared to using the electric field data.
Nevertheless, as demonstrated below, the analyses utilising both types of input data yield promising results for their prospective application in the analysis of particle data from kinetic simulations.

In the following sections, we present the results obtained for each type of the input data with the classification and regression algorithms performed by the CNNs and MLPs, and the AD algorithm conducted by CNNs. During the training process, the NN uses the validation dataset to compute the validation loss. If the loss remains stable over a certain number of iterations, indicating that the loss has reached a plateau, the training is terminated. In our NNs we use 30 epochs to assess the invariance of the validation loss. 
This is implemented to prevent unnecessary further training, which not only consumes additional time but, more importantly, guards against overfitting. Overfitting occurs when the NN excessively tunes itself to the training dataset, which results in a loss of generalization capability when dealing with previously unknown data. We divide the full dataset into training, validation and testing datasets as 119,700 (47.5\%), 69,300 (27.5\%) and 63,000 (25\%) particles, respectively. The parameters of NNs used in our analyses are listed in the \nameref{sec:appendix} appendix.

\subsection{Input data: \texorpdfstring{$p_x$}{px}, \texorpdfstring{$p_y$}{py}, \texorpdfstring{$p_z$}{pz} momentum components time series} \label{sec:results_momentum}

\subsubsection{Classification}\label{sec:momentum_class}
\begin{figure}[ht]
    \centering
    \begin{subfigure}{\columnwidth}
        \centering
        \begin{tabular}{c||c c c}
            \backslashbox{True}{Pred.}  & 0 & 1 & 2 \\ \hline \hline
            0 & 61440 & 487 & 0 \\ 
            1 & 1 & 930 & 24 \\
            2 & 0 & 7 & 111
        \end{tabular}
        \phantomcaption
        \label{fig:conf_cnn}
    \end{subfigure}
    \begin{tikzpicture}[overlay]
        \node at (-2.3,0.25) {(a)};
    \end{tikzpicture}
    \vspace{1em}
    \begin{subfigure}{\columnwidth}
        \centering
        \begin{tabular}{c c c c}
            \hline \hline
            Class & Precision & Recall & F1-score \\ \hline
            0 & $\simeq$1.000 & 0.992 & 0.996 \\
            1 & 0.653 & 0.974 & 0.782 \\
            2 & 0.822 & 0.941 & 0.878 \\
            \hline\hline
        \end{tabular}
        \phantomcaption
        \label{tab:report_cnn}
    \end{subfigure}
    \begin{tikzpicture}[overlay]
        \node at (-2.3,0.65) {(b)};
    \end{tikzpicture}
    \vspace{1em}
    \begin{subfigure}{\columnwidth}
        \begin{overpic}[width=\columnwidth]{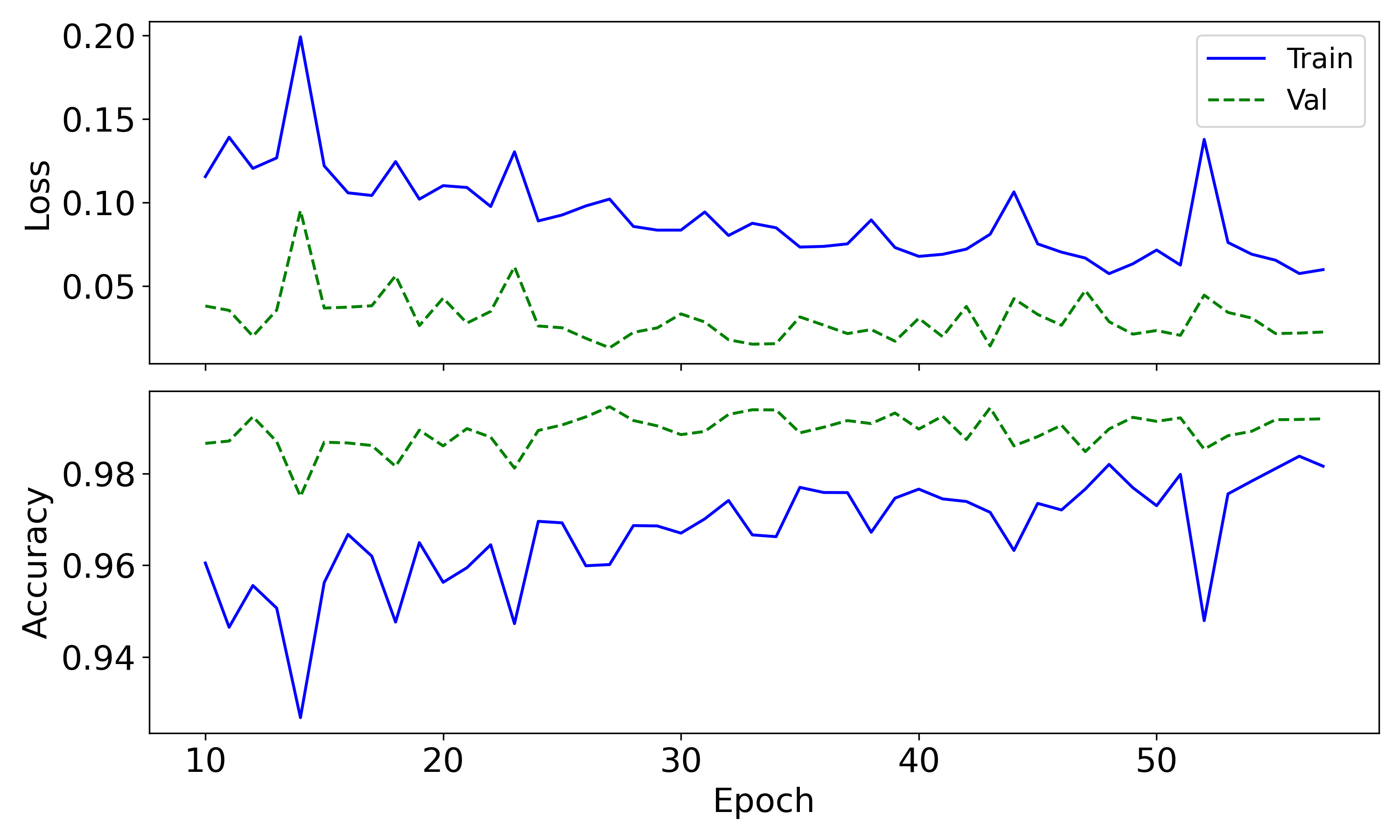}
            \setlength{\height}{\heightof{\includegraphics{Figures/cnn_cls/pxyz_cls_cnn_2_loss.png}}}
            \put(0.12\columnwidth,0.31\height){(c)}
        \end{overpic}
        \phantomcaption
        \label{fig:loss_cnn}
    \end{subfigure}
    \caption{Results for classification with CNN. Panel (a): the confusion matrix. The numbers show the distribution of data across all possible combinations of true and predicted labels for the 3 considered classes. Panel (b): precision $P$, recall, $R$, and $F1$-score derived from the confusion matrix in panel (a). Panel (c): the loss and accuracy over the epochs required to train and validate the NN.}
    \label{fig:results_cls_cnn}
\end{figure}

\begin{figure}[ht]
    \centering
    \begin{subfigure}{\columnwidth}
        \centering
        \begin{tabular}{c||ccc}
            \backslashbox{True}{Pred.}  & 0 & 1 & 2 \\ \hline\hline
            0 & 61697 & 230 & 0  \\
            1 & 36 & 895 & 24 \\
            2 & 0 & 36 & 82
        \end{tabular}
        \phantomcaption
        \label{fig:conf_mlp_cls}
    \end{subfigure}
    \begin{tikzpicture}[overlay]
        \node at (-2.3,0.25) {(a)};
    \end{tikzpicture}
    \vspace{1em}
    \begin{subfigure}{\columnwidth}
            \centering
            \begin{tabular}{c c c c}
                \hline\hline
                Class & Precision & Recall & F1-score \\ \hline
                0 & 0.999 & 0.996 & 0.998 \\
                1 & 0.771 & 0.937 & 0.846 \\
                2 & 0.774 & 0.695 & 0.732 \\
                \hline\hline
            \end{tabular}
            \phantomcaption
            \label{tab:report_mlp}
    \end{subfigure}
    \vspace{1em}
    \begin{tikzpicture}[overlay]
        \node at (-2.3,0.65) {(b)};
    \end{tikzpicture}
    \begin{subfigure}{\columnwidth}
        \begin{overpic}[width=\columnwidth]{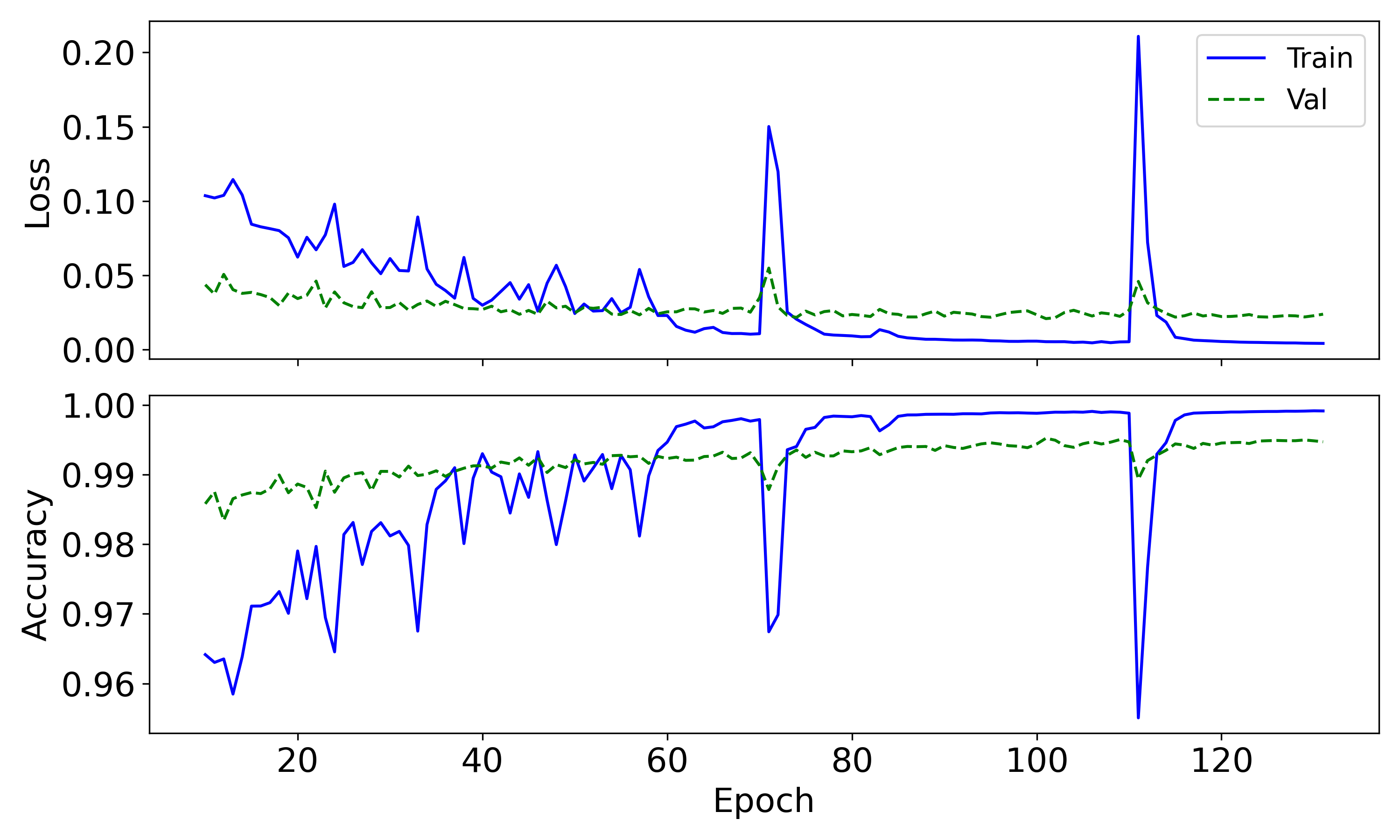}
            \setlength{\height}{\heightof{\includegraphics{Figures/mlp_cls/pxyz_cls_mlp_2_loss.png}}}
            \put(0.12\columnwidth,0.31\height){(c)}
        \end{overpic}
        \phantomcaption
        \label{fig:loss_mlp_cls}
    \end{subfigure}
    \caption{Results for classification with MLP (compare~Fig.~\ref{fig:results_cls_cnn}).}
    \label{fig:results_cls_mlp}
\end{figure}

Figures \ref{fig:results_cls_cnn} and \ref{fig:results_cls_mlp} show the confusion matrix, the progression of loss and accuracy across training and validation epochs for NNs, along with a table presenting the precision, recall, and F1-score of the three classes
defined in Section~\ref{sec:cls}. These metrics are provided
for CNN and MLP models, respectively. 
The confusion matrices are used here to evaluate the performance of our classification models, as they compare predicted classes with respect to the true classes. 
The best performance is achieved with a mostly diagonal matrix. The performance metrics -- precision $P$, recall $R$, and $F1$-score -- are calculated from the confusion matrix, $M$, as follows:
\begin{equation}
    P_i = \frac{M_{ii}}{\sum_jM_{ij}}, \quad
    R_j = \frac{M_{jj}}{\sum_iM_{ij}}, \quad
    F1_i = 2\frac{P_i\cdot R_i}{P_i+R_i},
\end{equation}
where $i$ and $j$ represent predicted (columns) and true (rows) values, respectively.
The precision metric quantifies the accuracy of positive predictions made by the model, the recall, known also as sensitivity, measures the ratio of true positive predictions to all actual positive instances, and the $F1$-score combines both $P$ and $R$. The use of a harmonic mean in the definition of the $F1$-score is a particularly useful metric for our imbalanced dataset, as it can be high only if both $P$ and $R$ are high.

One can see in Figures~\ref{fig:loss_cnn} and \ref{fig:loss_mlp_cls} that it takes approximately 60-70 epochs for the loss to stabilize during the training and validation phases of the CNN and MLP networks with the classification algorithm.
Note the spikes visible in the loss and accuracy figures. The training dataset is divided into small batches, each containing a random subset of the training data. This subset changes every epoch. Occasionally, batches may contain unfavorable values, such as batches consisting entirely of class 0 particles, which can slightly affect the results and lead to deviations in the weights of the NN, resulting in a higher loss than expected. However, the NN optimiser recognizes this temporary increase in loss and returns to a low loss after a small number of epochs. 
This behavior can be clearly observed in Figure~\ref{fig:loss_mlp_cls}, e.g., between epochs $\sim 70$ and $\sim75$, and epochs $\sim110$ and $\sim115$.

It is also worth noting that sometimes the validation loss (accuracy) is lower (higher) than the training loss (accuracy) e.g., Figure~\ref{fig:loss_cnn}). As shown in Figure 2, our dataset is heavily imbalanced. Since the training and validation datasets are randomly sampled from the entire dataset (with no shared data), it is possible for the validation set to contain particularly favourable data points, which can lead to a higher accuracy compared to the training set. However, as long as both the training and validation accuracies are increasing over the course of the epochs, we can conclude that the neural network has been successfully trained. This is evident from the results shown in Figures~\ref{fig:loss_cnn} and~\ref{fig:loss_mlp_cls}.

As shown in Tables \ref{tab:report_cnn} and \ref{tab:report_mlp}, class 0, being the most abundant, achieves a very high F1-score, as expected. Both class 1 and class 2 also achieve high F1-scores. Interestingly, class 1 has a lower F1-score than class 2, despite containing fewer particles. This can be attributed to the distinctive time series shapes of the very high-energy particles in class 2 (characterized by large-amplitude momentum variations, see Fig.~\ref{fig:bun_traj}d-f), which aid the NN in their recognition. For both NNs, the results are notably robust, with F1-scores exceeding 0.7 for all classes. Overall, both the CNN and MLP demonstrate remarkably good performance.

\subsubsection{Regression} \label{sec:results_reg}
\begin{figure}
    \centering
    \begin{subfigure}{\columnwidth}
        \begin{overpic}[width=\columnwidth]{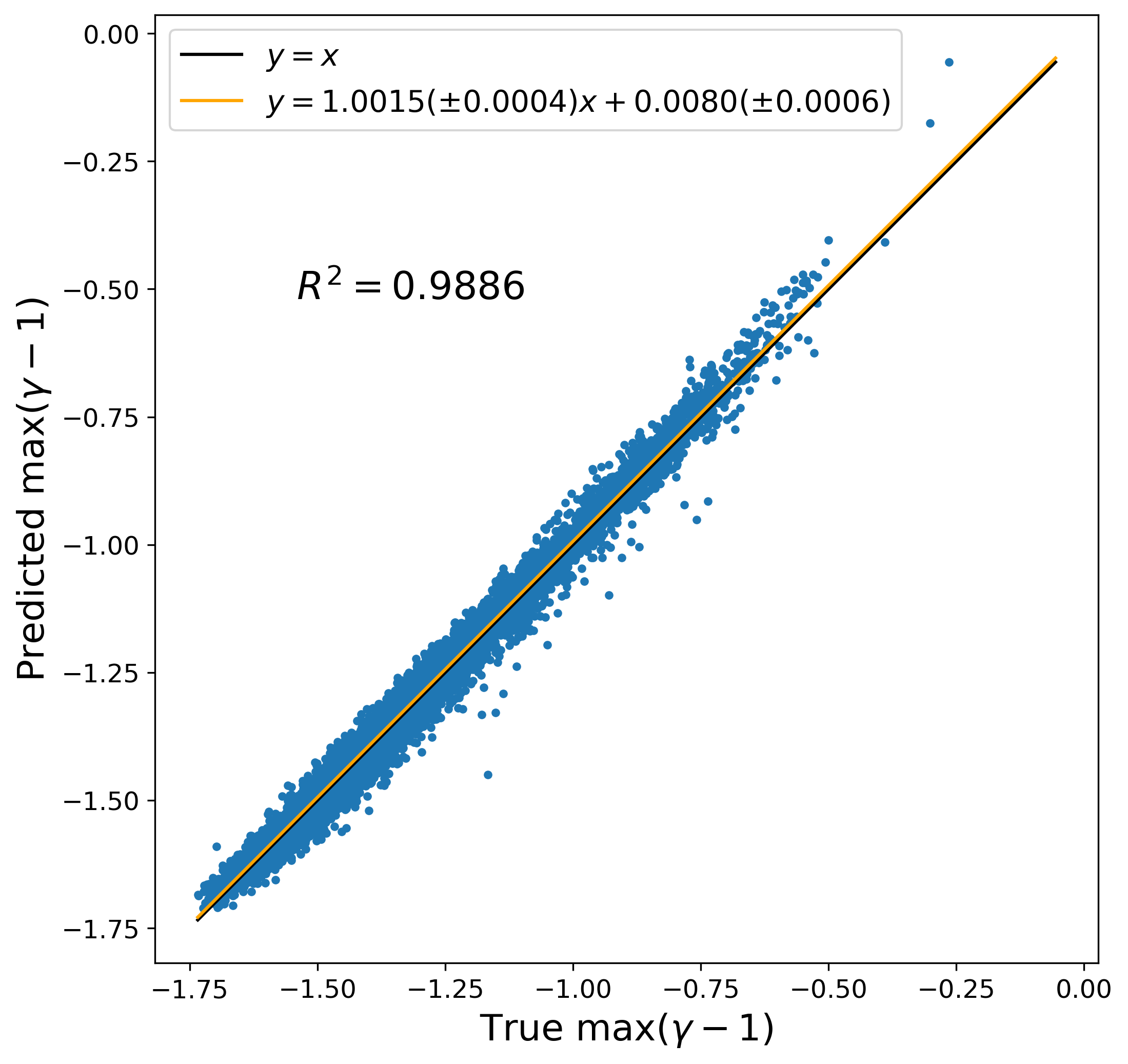}
            \setlength{\height}{\heightof{\includegraphics{Figures/cnn_reg/pxyz_reg_cnn_2_regression.png}}}
            \put(0.15\columnwidth,0.39\height){(a)}
        \end{overpic}
        \phantomcaption
        \label{fig:plot_cnn}
    \end{subfigure}
    \begin{subfigure}{\columnwidth}
        \begin{overpic}[width=\columnwidth]{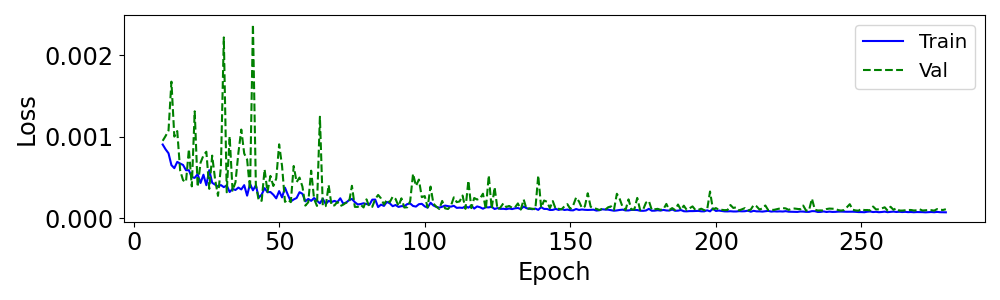}
            \setlength{\height}{\heightof{\includegraphics{Figures/cnn_reg/pxyz_reg_cnn_2_loss.png}}}
            \put(0.13\columnwidth,0.28\height){(b)}
        \end{overpic}
        \phantomcaption
        \label{fig:loss_cnn_reg}
    \end{subfigure}
    \begin{subfigure}{\columnwidth}
        \begin{overpic}[width=\columnwidth]{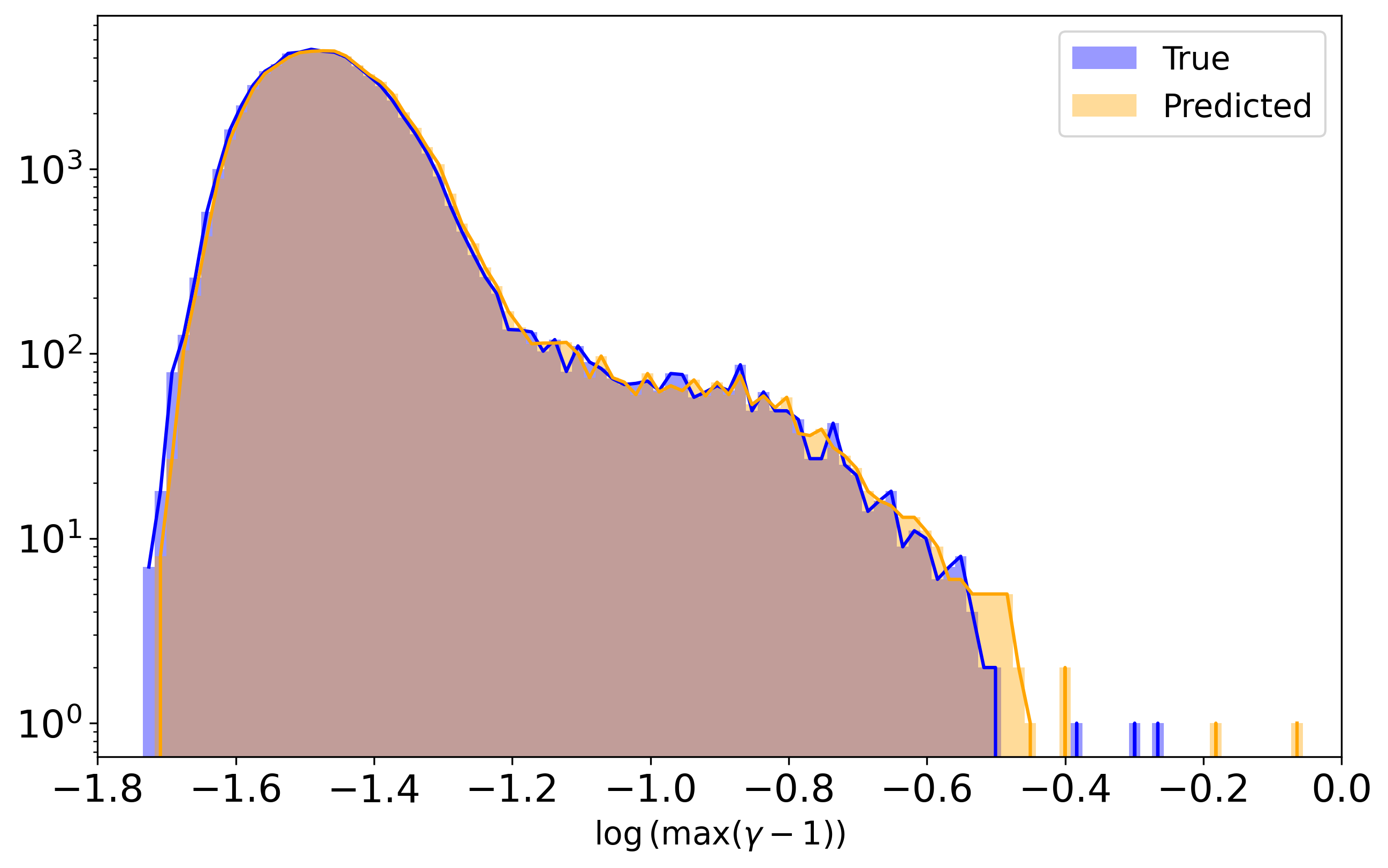}
            \setlength{\height}{\heightof{\includegraphics{Figures/cnn_reg/pxyz_reg_cnn_2_histogram.png}}}
            \put(0.08\columnwidth,0.36\height){(c)}
        \end{overpic}
        \phantomcaption
        \label{fig:hist_cnn}
    \end{subfigure}
    \caption{Results for regression with CNN. Panel (a:) the regression of the true value of the energy and the energy predicted by the NN.
    Panel (b): the evolution of loss during training.
    Panel (c): comparison of the training values and the predicted values during the testing phase.
    }
    \label{fig:results_reg_cnn}
\end{figure}

\begin{figure}
    \centering
    \begin{subfigure}{\columnwidth}
        \begin{overpic}[width=\columnwidth]{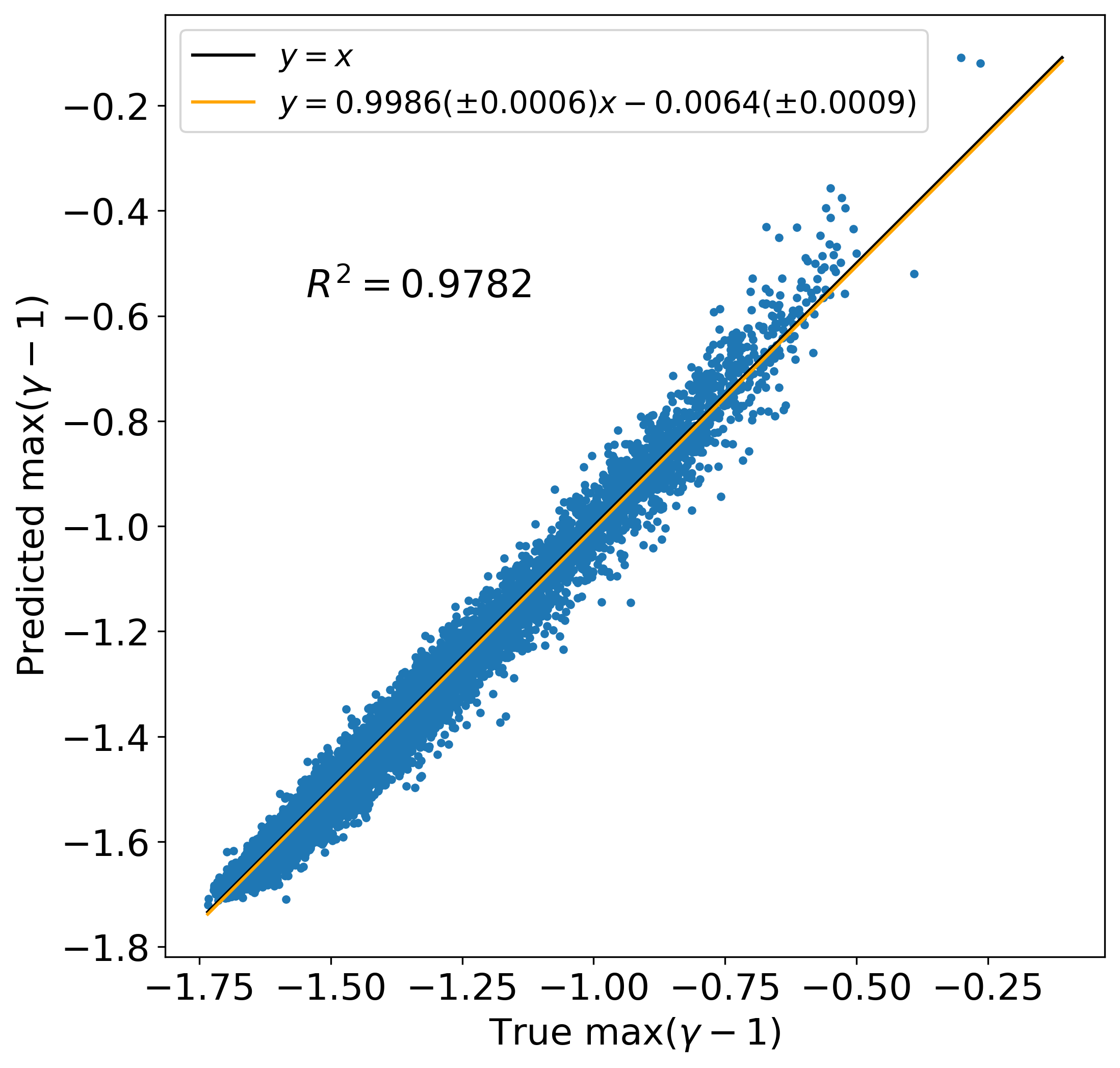}
            \setlength{\height}{\heightof{\includegraphics{Figures/mlp_reg/pxyz_reg_mlp_2_regression.png}}}
            \put(0.16\columnwidth,0.39\height){(a)}
        \end{overpic}
        \phantomcaption
        \label{fig:plot_mlp}
    \end{subfigure}
    \begin{subfigure}{\columnwidth}
        \begin{overpic}[width=\columnwidth]{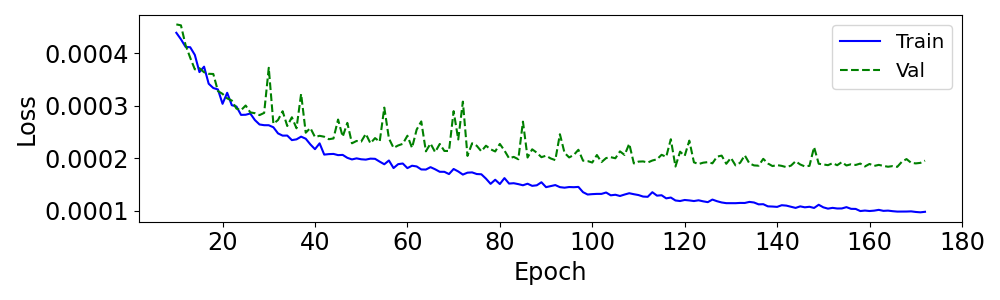}
            \setlength{\height}{\heightof{\includegraphics{Figures/mlp_reg/pxyz_reg_mlp_2_loss.png}}}
            \put(0.15\columnwidth,0.28\height){(b)}
        \end{overpic}
        \phantomcaption
        \label{fig:loss_mlp_reg}
    \end{subfigure}
    \begin{subfigure}{\columnwidth}
        \begin{overpic}[width=\columnwidth]{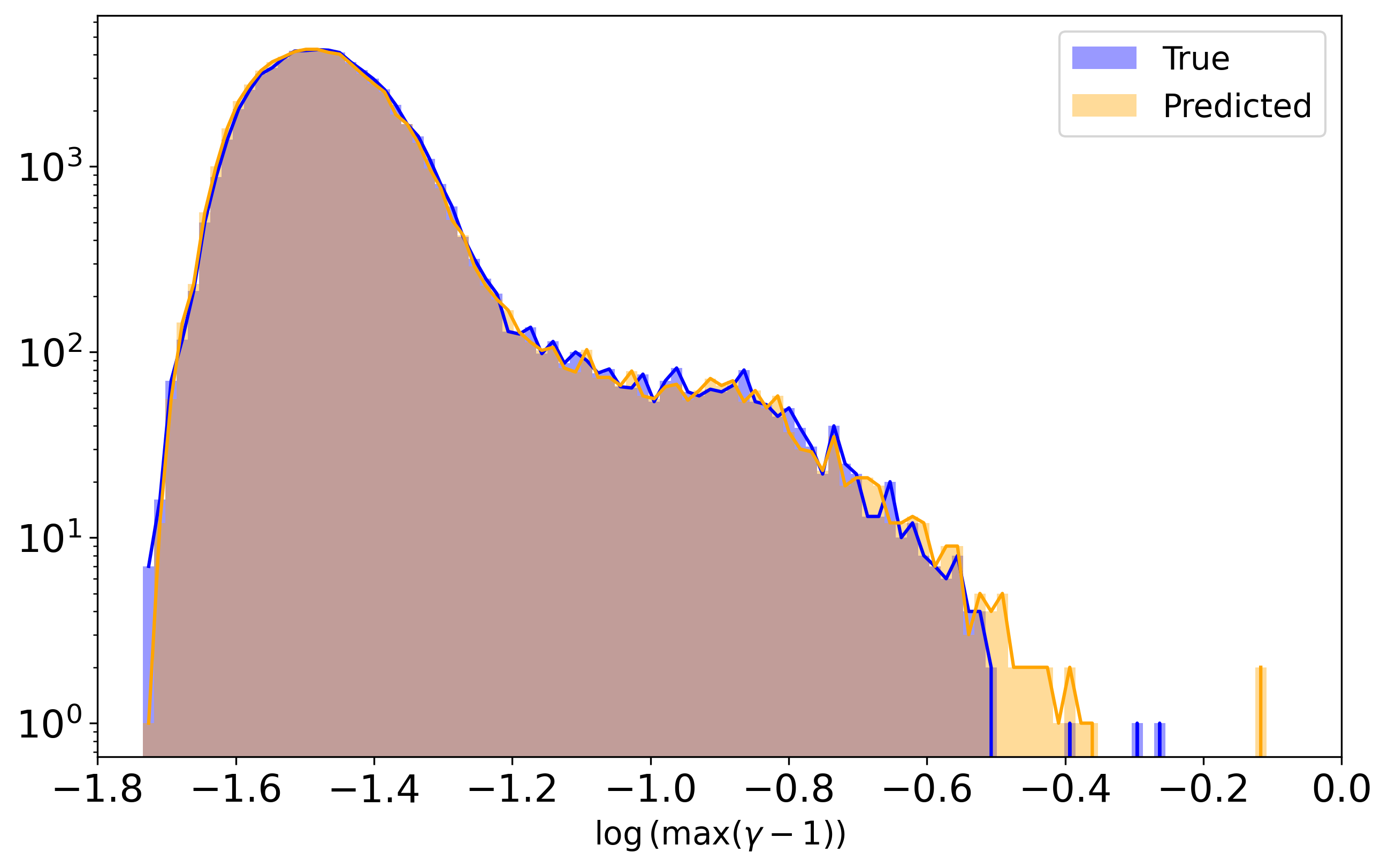}
            \setlength{\height}{\heightof{\includegraphics{Figures/mlp_reg/pxyz_reg_mlp_2_histogram.png}}}
            \put(0.08\columnwidth,0.36\height){(c)}
        \end{overpic}
        \phantomcaption
        \label{fig:hist_mlp}
    \end{subfigure}
    \caption{Results for regression with MLP (compare Fig.~\ref{fig:results_reg_cnn}).}
    \label{fig:results_reg_mlp}
\end{figure}

Figures~\ref{fig:results_reg_cnn} and \ref{fig:results_reg_mlp} show for the CNN and MLP, respectively, the linear regression applied to the true ($x$-axis) and predicted ($y$-axis) values of the maximum kinetic energy in panels (a), the loss obtained during training in panels (b), and the histogram of both true and predicted values in panels (c). The linear regression fits are shown in panels (a) with orange lines, whereas the black lines $y=x$ represent a perfect score. 
In our calculations, we use the Huber loss function, which yields the lowest loss among the loss functions we have tested.

For both the CNN and MLP, the regression lines lie very close to the black line, as desired. To qualitatively assess the results, the coefficient of determination $R^2$ is computed. We obtain $R^2=0.9886$ for the CNN and $R^2=0.9782$ for the MLP, indicating near-perfect fits. In both cases we observe that  high-energy particles are slightly overpredicted, showing higher energies than they should.
These effects are visible in the histograms of true (blue) and predicted (orange) values. For the CNN, there is a slight overprediction at high energies, but overall, the predicted histogram closely matches the true one. The MLP also shows a similar slight overprediction on the high-energy side.
One might expect the opposite, with particles being underpredicted due to the large number of low-energy particles on which the NNs are trained. However, it is possible that the strong effect of sample weighting on high-energy particles biases the regression towards these particles.
In conclusion, both the CNN and MLP yield very similar goodness-of-fit, indicating that either could be used for particle data analysis. We note however, that the MLP performs slightly faster.

\subsubsection{Anomaly Detection}
In this section, we present results for the autoencoder made of convolutional layers. While the MLP-based alternative was also tested, its results were significantly inferior, leading us to focus solely on the CNN-based autoencoder.

As the first approach, the autoencoder was trained exclusively with low-energy particles, with $\max{(\gamma-1)}<0.07$, to prevent exposure to any uncommon data that could bias the results towards high-energy particles, thereby potentially deteriorating the outcome significantly. This approach to anomaly detection is called semi-supervised anomaly detection, 
as we can confidently identify a portion of our dataset composed of common particles only, i.e. low-energy thermal population. Higher-energy particles are excluded from this sample, including those within the energy range $\max{(\gamma-1)}\approx 0.07-0.1$, in which the thermal and non-thermal populations blend together. As the second approach, the autoencoder underwent training using the entire training dataset, including the non-thermal population, to investigate the effect of uncommon input data during the training phase.

\begin{figure*}[ht]
    \centering
    \begin{subfigure}{.49\textwidth}
        \begin{overpic}[width=\linewidth]{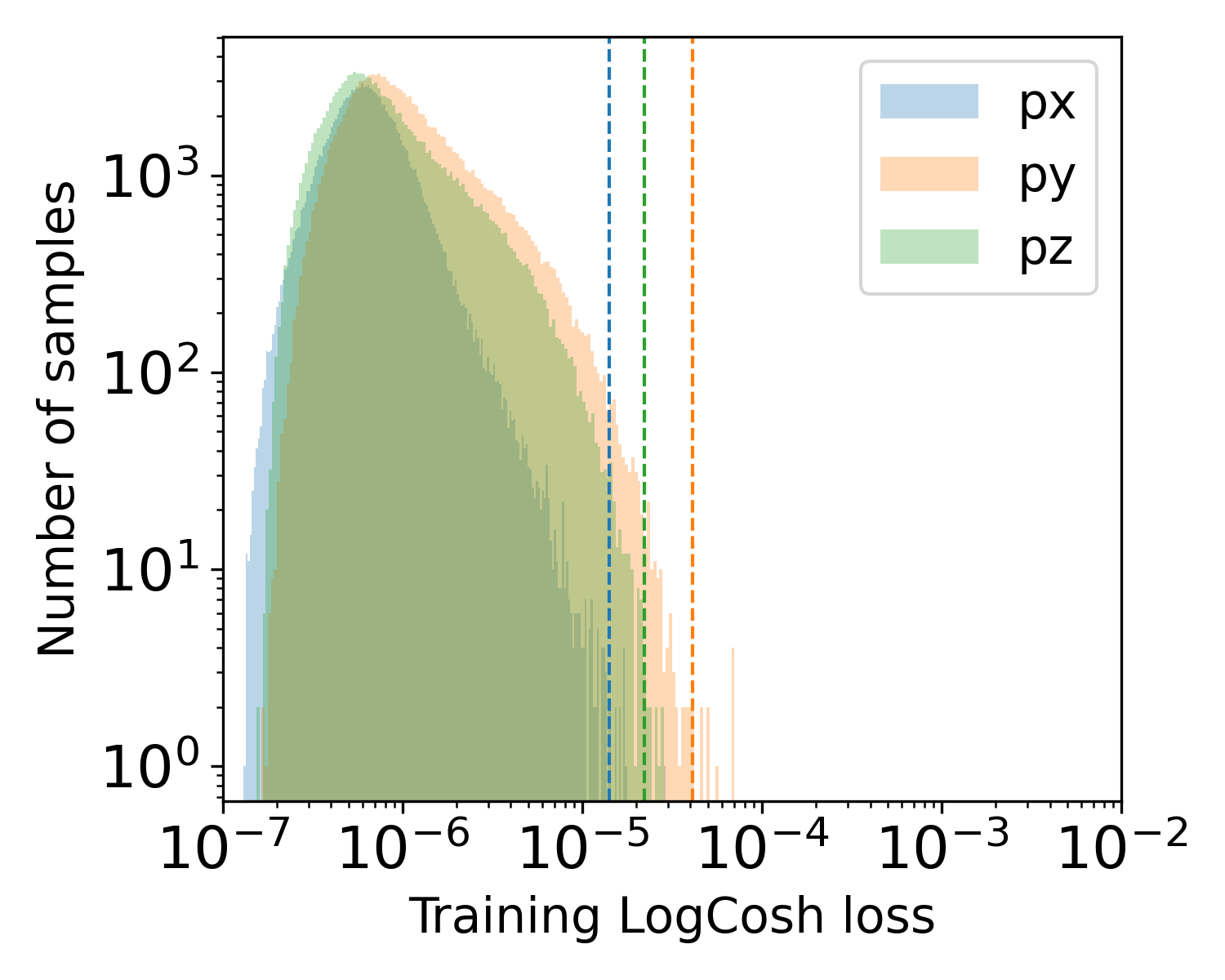}
            \setlength{\height}{\heightof{\includegraphics{Figures/ad/pxyz_ad_cnn_007_10_training_histogram.png}}}
            \put(0.19\linewidth,0.63\height){(a)}
        \end{overpic}
        \phantomcaption
        \label{fig:mae_loss}
    \end{subfigure}
    \begin{subfigure}{.49\textwidth}
        \begin{overpic}[width=\linewidth]{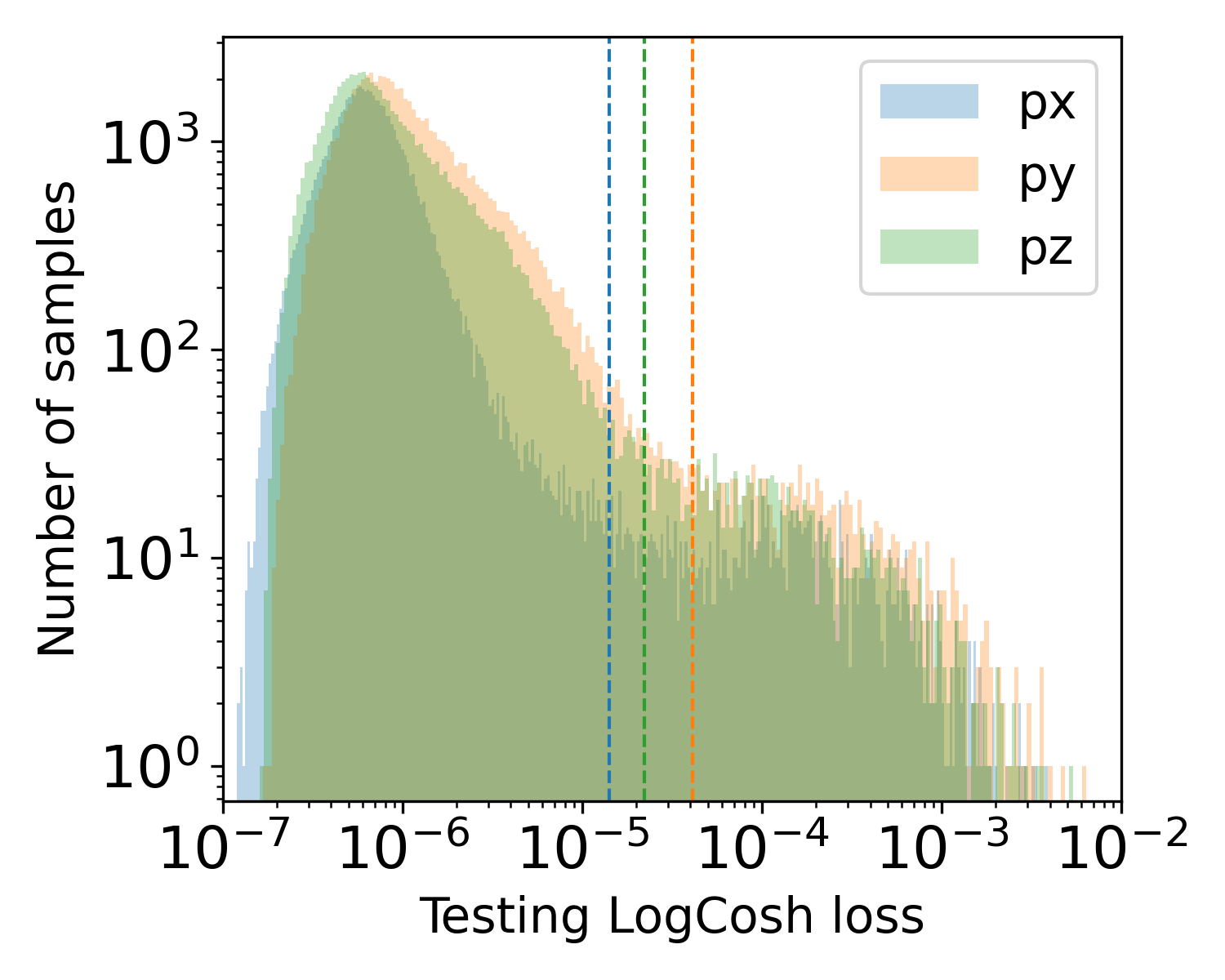}
            \setlength{\height}{\heightof{\includegraphics{Figures/ad/pxyz_ad_cnn_007_10_anomalies_histogram.png}}}
            \put(0.19\linewidth,0.63\height){(b)}
        \end{overpic}
        \phantomcaption
        \label{fig:anomalies_loss}
    \end{subfigure}

    \begin{subfigure}{1\textwidth}
        \begin{overpic}[width=\linewidth]{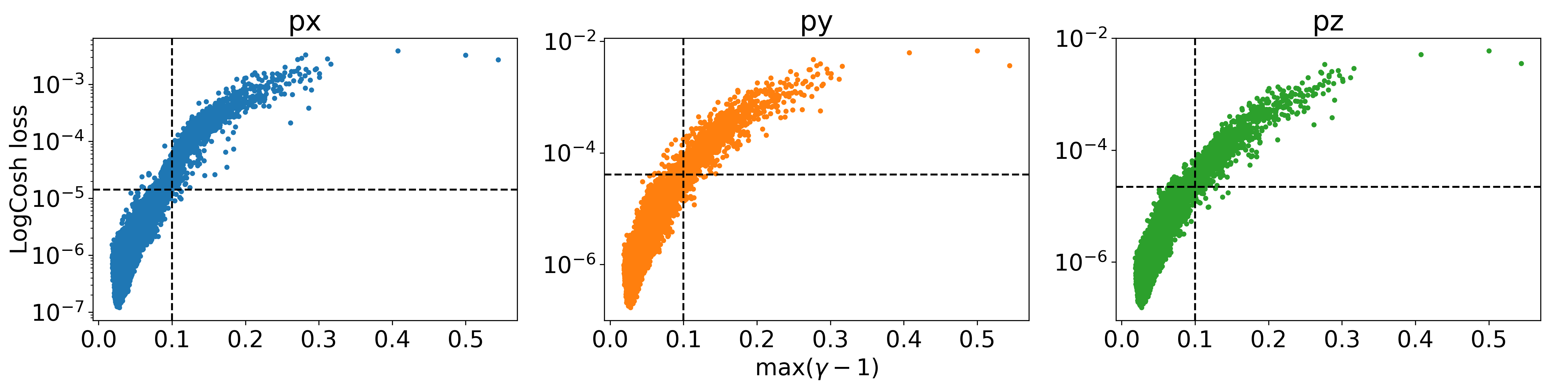}
            \setlength{\height}{\heightof{\includegraphics{Figures/ad/pxyz_ad_cnn_007_10_anomaly_detection_label_vs_loss.png}}}
            \put(0.065\linewidth,0.34\height){(c)}
        \end{overpic}
        \phantomcaption
        \label{fig:label_vs_loss}
    \end{subfigure}

    \begin{subfigure}{\textwidth}
            \centering
            \begin{tabular}{c|cc}
                 & Predicted Non-anomaly & Predicted Anomaly\\ \hline
                Actual Non-anomaly & 61614 & 313 \\
                  Actual Anomaly & 4 & 1069 \\
            \end{tabular}
            \phantomcaption
            \label{tab:ad_matrix}
            \begin{tikzpicture}[overlay]
                \node at (-0.45\textwidth,0.6) {(d)};
            \end{tikzpicture}
    \end{subfigure}
    \caption{Results for semi-supervised anomaly detection with CNN. Panels (a) and (b): the training and testing loss for $p_x$, $p_y$ and $p_z$, respectively. The dotted vertical lines indicate the thresholds used for AD, which during testing, split the histogram into anomalies and non-anomalies. Panel (c): the maximum kinetic energy vs. the testing loss. The horizontal dashed lines represents thresholds for each momentum component and the vertical lines depict our division between energetic and non-energetic particles. 
    Panel (d): the confusion matrix obtained from results in panel (c).   
    }
    \label{fig:results_ad}
\end{figure*}

Figure~\ref{fig:mae_loss} shows the loss obtained after training the NN with low-energy particles only. We choose the thresholds for each momentum component to be the tenth highest loss calculated, as indicated with the dashed lines in Figure~\ref{fig:mae_loss}. Although we could have chosen the highest loss as the threshold, the NN training is not perfect, and some particles may produce excessively high losses,  which could bias the testing results and lead to worse overall predictions. Therefore, we introduce some flexibility by selecting a slightly lower value. 
Figure~\ref{fig:anomalies_loss} illustrates the testing loss. If the loss for an input variable exceeds the threshold, it is considered an anomaly. Since particles are characterized by three momentum components, we consider a particle an anomaly if the loss in any of the momentum components is higher than the defined threshold. 

Figure~\ref{fig:label_vs_loss} summarises our results. In each panel, the horizontal dashed lines represent the thresholds computed for each feature, while the vertical lines at $\max{(\gamma-1)}<0.1$ indicate our separation between energetic and non-energetic (thermal) particles, identical to the one used in classification to distinguish class 0 from class 1 and 2 particles (see Table \ref{tab:classes}).
Ideally, all energetic particles would lie in the top-right quadrant. However, time series of energetic particles located close to the crossing point where both dashed lines intersect become similar to those of non-energetic particles.
An energised particle may therefore be mistaken for a thermal one, leading the autoencoder to fail in differentiating them, often labeling these ambiguous time series as non-anomalies. 
The results from Figure~\ref{fig:label_vs_loss} are summarised in the confusion matrix in Figure~\ref{tab:ad_matrix}.
Out of 1073 energetic particles (actual anomalies) in our testing sample of 63,000 particles, the autoencoder correctly predicts 1069 particles, yielding a superb sensitivity (recall) of 0.996. However, at the same time the autoencoder predicts more anomalies than it should have, with 313 non-energetic particles classified as anomalies, resulting in the precision of 0.774, which is still remarkable. These misclassified non-energetic particles likely reside very close to the vertical dashed line (energy separation), meaning that their time series are somewhat similar to those of energetic particles. Overall, AD with CNN-based autoencoder performs remarkably well with the accuracy (the ratio of correctly predicted observations to the total observations) of 99.6\% and total F1-score of 87.9\%. 

This result can be elucidated with an example: if we trace new particles and feed them to the autoencoder, we can be confident that more than 3/4 of the anomalies (approximately 77\%) are pre-accelerated particles, and we miss less than 0.5\% of energetic particles.

\begin{figure*}
    \centering
    \begin{subfigure}{0.49\textwidth}
        \includegraphics[width=\linewidth]{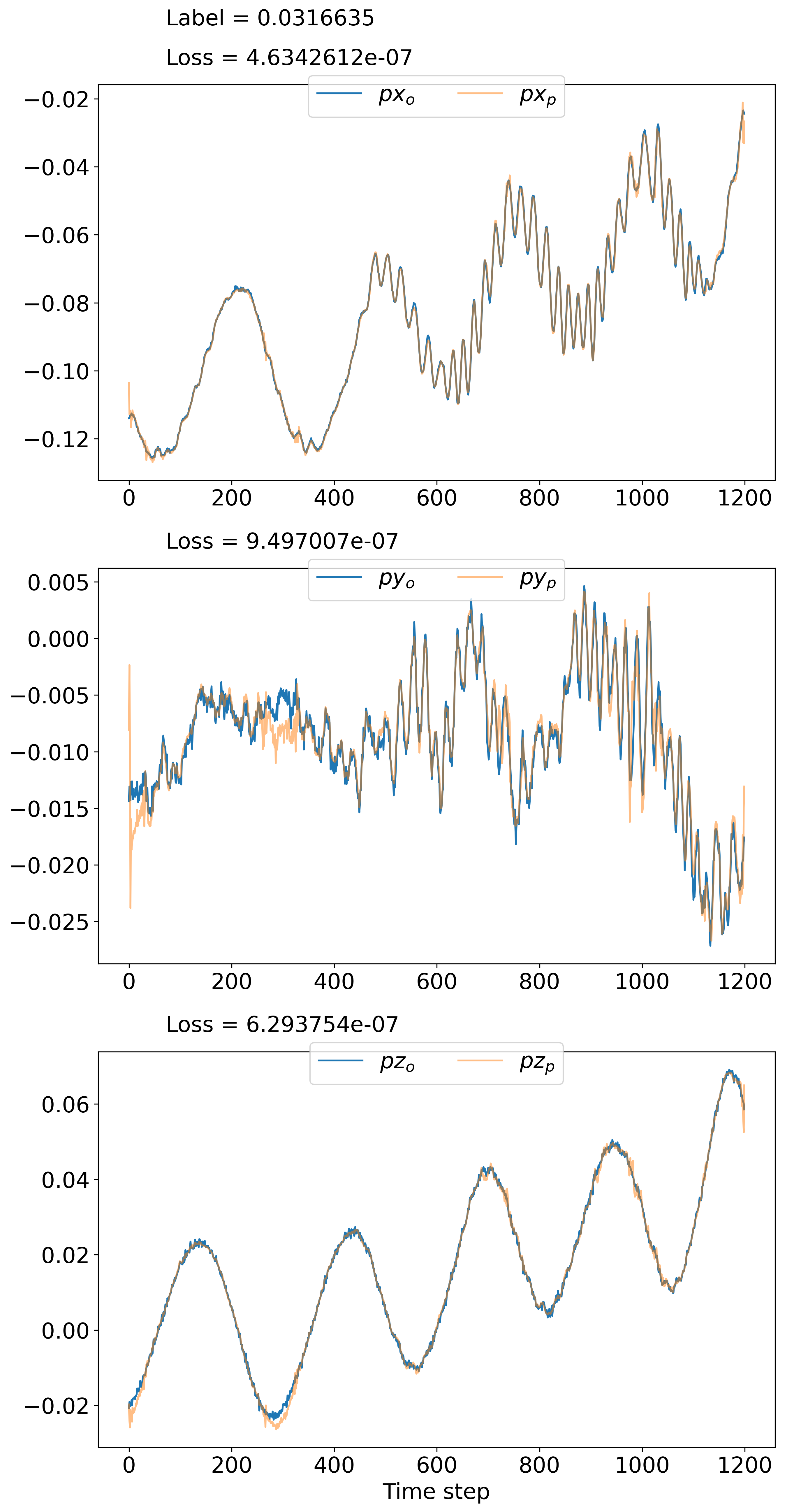}
        \phantomcaption
        \label{fig:non_anomaly}
    \end{subfigure}
    \hfill
    \begin{subfigure}{0.49\textwidth}
        \includegraphics[width=\linewidth]{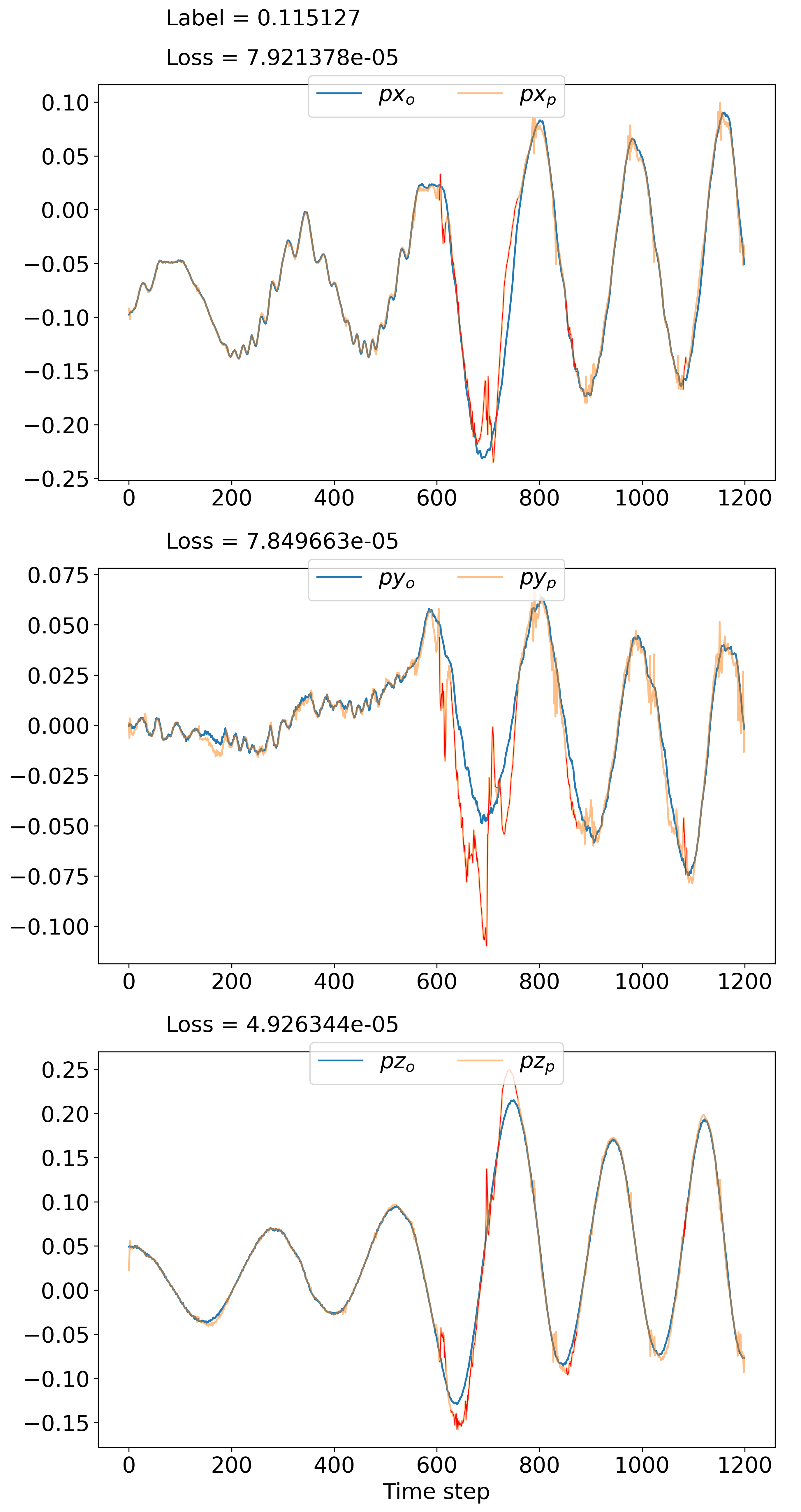}
        \phantomcaption
        \label{fig:anomaly}
    \end{subfigure}
    \caption{Momentum components time series and results of anomaly detection for two example particles: a non-energetic electron (a non-anomaly) with the maximum kinetic energy $y=0.0316635$ in \emph{left} panels, and an energetic electron (an anomaly) with $y=0.115127$ in \emph{right} panels. The original and predicted time series are shown with blue and orange lines, respectively. The red lines in right panels indicate anomalous sections where the predictions differ most from the original time series. Losses indicated at each panel are to be compared with thresholds shown in Fig.~\ref{fig:mae_loss}.}
    \label{fig:ad_examples}
\end{figure*}

Figure~\ref{fig:ad_examples} shows two examples of how the autoencoder detects anomalies. In the \emph{left} panels, the predictions of the momentum evolution (orange lines) match quite well the original time series (blue lines) for all three components, resulting in losses much lower than the thresholds (compare Fig.~\ref{fig:mae_loss}). Therefore, this particle is classified as a non-anomaly, which is correct since it has the maximum kinetic energy $y\simeq 0.0317$. In contrast, the predicted and original time series for the particle presented in the \emph{right} panels show differences, highlighted in red. These differences are least significant for the $p_z$ momentum component, but deviations in all three components result in losses higher than the respective thresholds. Therefore, this particle, with $y\simeq 0.115$, is correctly identified as an anomaly.

\begin{figure*}[ht]
    \centering
    \begin{subfigure}{.49\textwidth}
        \begin{overpic}[width=\linewidth]{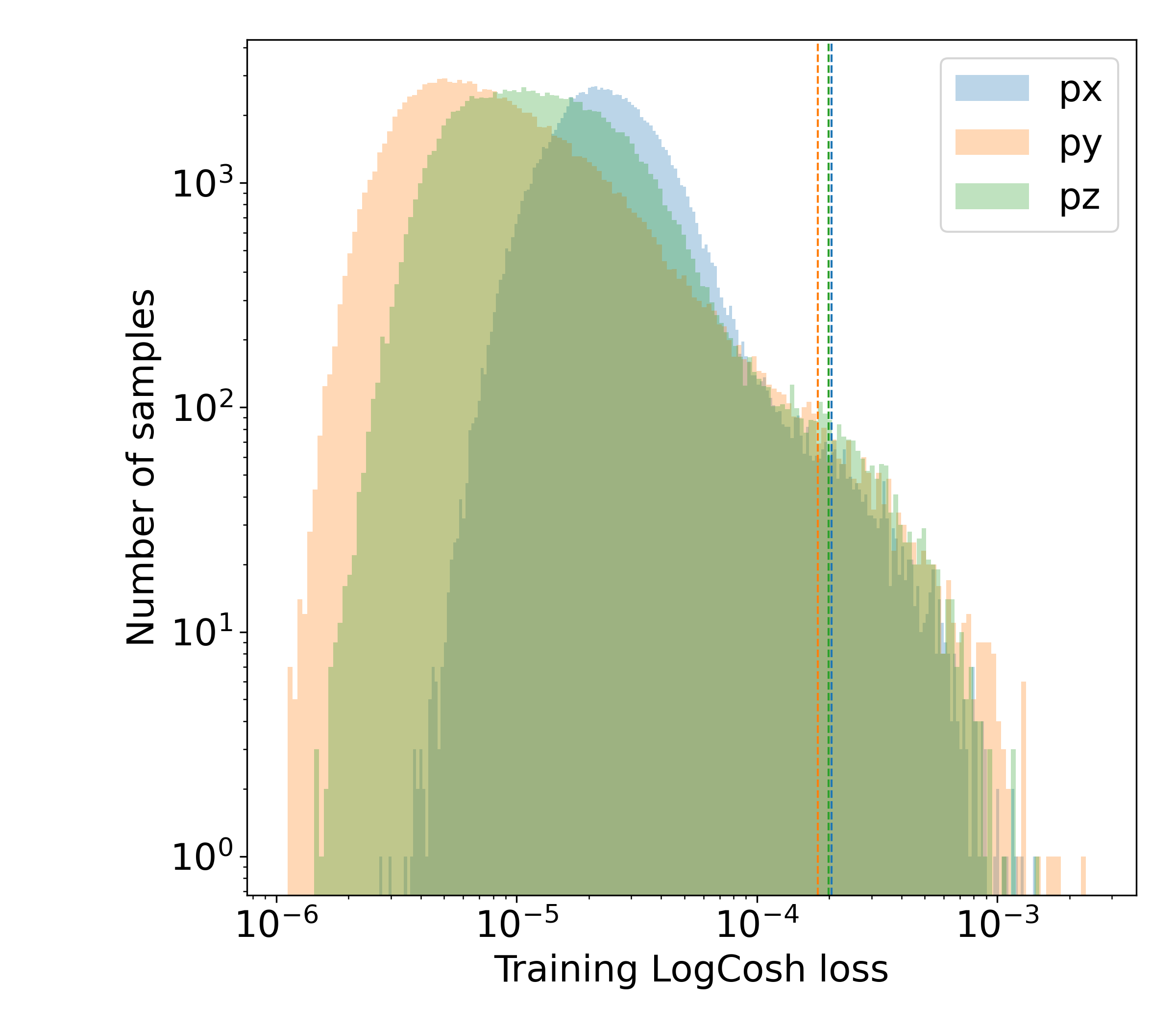}
            \setlength{\height}{\heightof{\includegraphics{Figures/ad_full/pxyz_full_ad_13_AD_LogCosh_loss.png}}}
            \put(0.22\linewidth,0.39\height){(a)}
        \end{overpic}
        \phantomcaption
        \label{fig:full_mae_loss}
    \end{subfigure}
    \begin{subfigure}{.49\textwidth}
        \begin{overpic}[width=\linewidth]{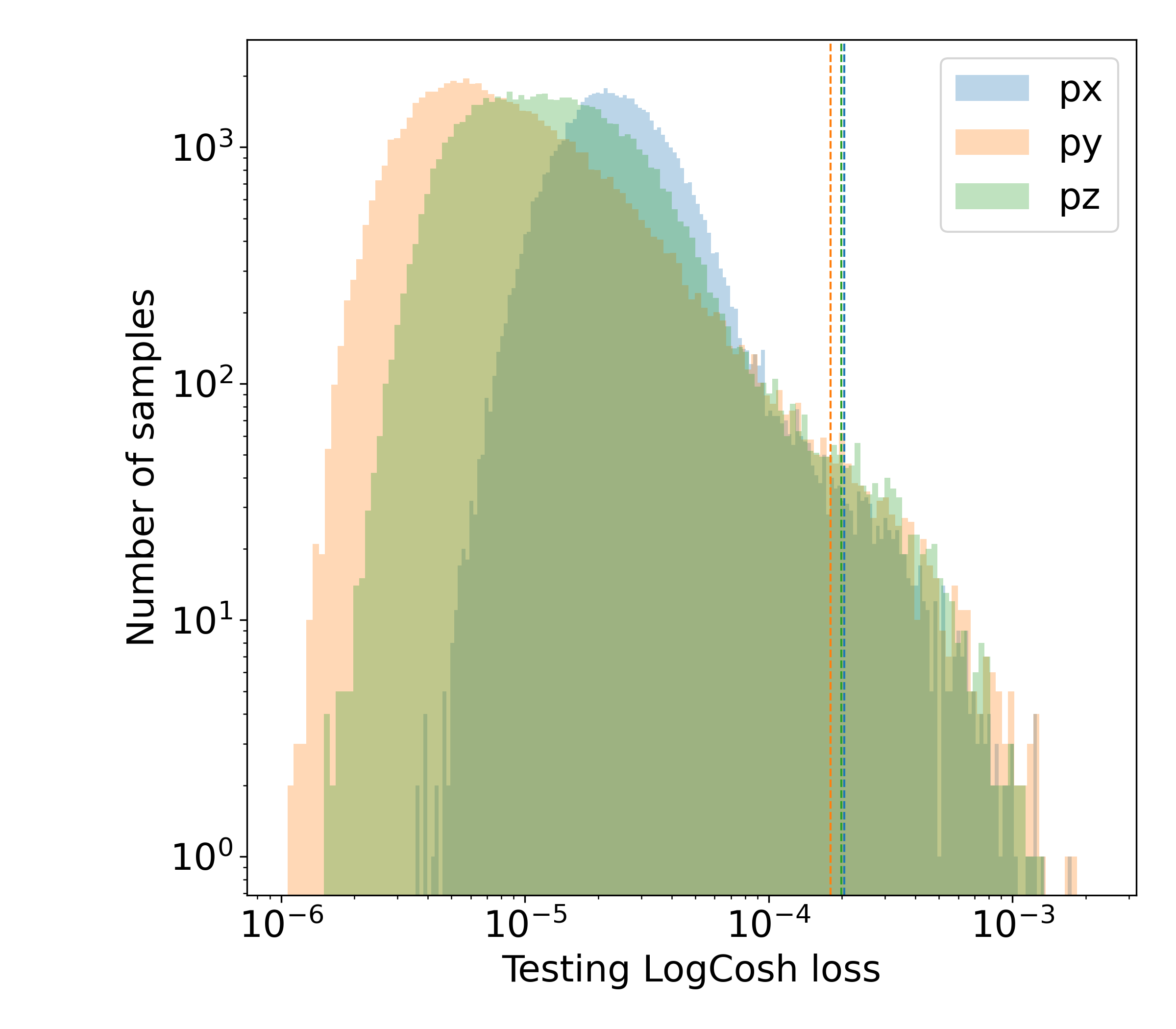}
            \setlength{\height}{\heightof{\includegraphics{Figures/ad_full/pxyz_full_ad_13_anomalies_loss.png}}}
            \put(0.22\linewidth,0.39\height){(b)}
        \end{overpic}
        \phantomcaption
        \label{fig:full_anomalies_loss}
    \end{subfigure}

    \begin{subfigure}{1\textwidth}
        \begin{overpic}[width=\linewidth]{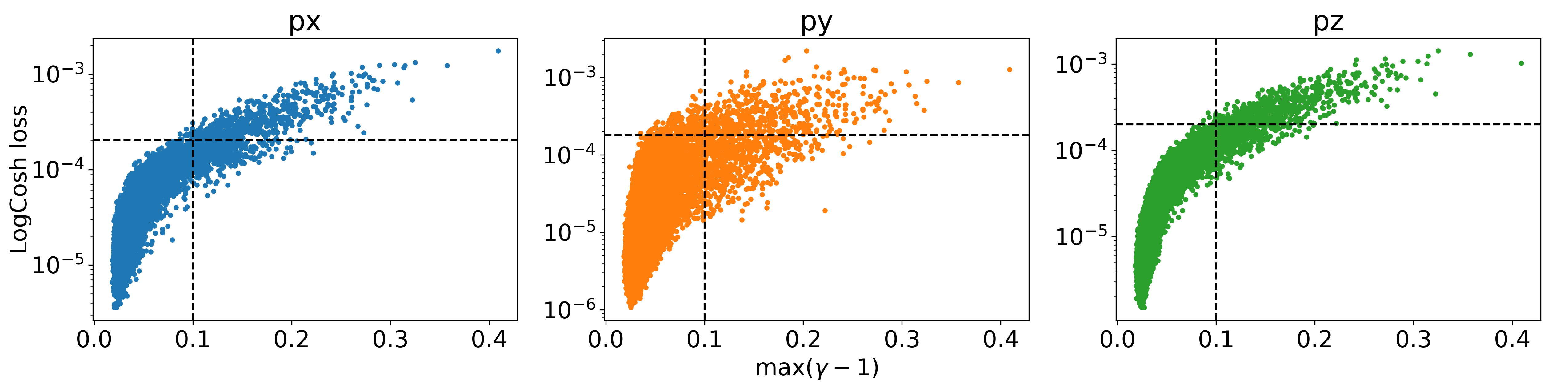}
            \setlength{\height}{\heightof{\includegraphics{Figures/ad_full/pxyz_full_ad_13_anomaly_detection_label_vs_loss.png}}}
            \put(0.065\linewidth,0.34\height){(c)}
        \end{overpic}
        \phantomcaption
        \label{fig:full_label_vs_loss}
    \end{subfigure}

    \begin{subfigure}{\textwidth}
            \centering
            \begin{tabular}{c|c c}
                 & Predicted Non-anomaly & Predicted Anomaly\\ \hline
                Actual Non-anomaly & 61932 & 13 \\
                  Actual Anomaly & 448 & 607 \\
            \end{tabular}
            \phantomcaption
            \label{tab:full_ad_matrix}
            \begin{tikzpicture}[overlay]
                \node at (-0.45\textwidth,0.6) {(d)};
            \end{tikzpicture}
    \end{subfigure}
    \caption{Results for fully unsupervised anomaly detection with CNN (compare Fig.~\ref{fig:results_ad}).}
    \label{fig:results_ad_full}
\end{figure*}

Figure~\ref{fig:results_ad_full} shows results for anomaly detection in which the NN was trained using particles selected randomly from the entire dataset, i.e., including supra-thermal and high-energy populations. The thresholds have been chosen to correspond to 1\% of the dataset. One can see in Figure~\ref{fig:full_mae_loss}, that the thresholds (dashed lines) are now positioned in the middle of the histogram rather than on the right, as in the previous case of AD. We have checked that higher or lower thresholds result in worse performance of the NN.
As expected, including the energetic populations in the training dataset degrades the overall performance.
Figure~\ref{fig:full_label_vs_loss} clearly shows a wider spread in all momentum components compared to Figure~\ref{fig:label_vs_loss},
which is largest for the $p_y$ component. To address this, we redefined the anomaly criterion to include two variable losses above the threshold instead of just one, as in the semi-AD case. The final result is shown in the confusion matrix in Figure~\ref{tab:full_ad_matrix}. Although the precision is high, $P=97.9\%$, as we obtain a small number of false anomalies (13), we are missing nearly 42\% of energetic particles (448), which results in low recall of 57.5\% and F1-score of 72.5\%. The AD performance is thus clearly better when only low-energy particles are used for NN training. 

\subsection{Input data: \texorpdfstring{$E_x$}{Ex} electric field} \label{sec:results_efield}
\subsubsection{Classification}
\begin{figure}[ht]
    \centering
    \begin{subfigure}{\columnwidth}
        \centering
        \begin{tabular}{c||ccc}
            \backslashbox{True}{Pred.}  & 0 & 1 & 2 \\ \hline\hline
            0 & 61685 & 242 & 0 \\
            1 & 121 & 800 & 34 \\
            2 & 0 & 56 & 62
        \end{tabular}
        \phantomcaption
        \label{fig:ex_cls_matrix}
    \end{subfigure}
    \begin{tikzpicture}[overlay]
        \node at (-2.3,0.25) {(a)};
    \end{tikzpicture}
    \vspace{1em}
    \begin{subfigure}{\columnwidth}
        \centering
        \begin{tabular}{cccc}
            \hline\hline
            Class & Precision & Recall & F1-score \\ \hline
            0 & 0.998 & 0.996 & 0.997 \\
            1 & 0.729 & 0.838 & 0.779 \\
            2 & 0.646 & 0.525 & 0.579 \\
            \hline\hline
        \end{tabular}
        \phantomcaption
        \label{tab:ex_report_cnn}
    \end{subfigure}
    \begin{tikzpicture}[overlay]
        \node at (-2.3,0.65) {(b)};
    \end{tikzpicture}
    \begin{subfigure}{\columnwidth}
        \begin{overpic}[width=\columnwidth]{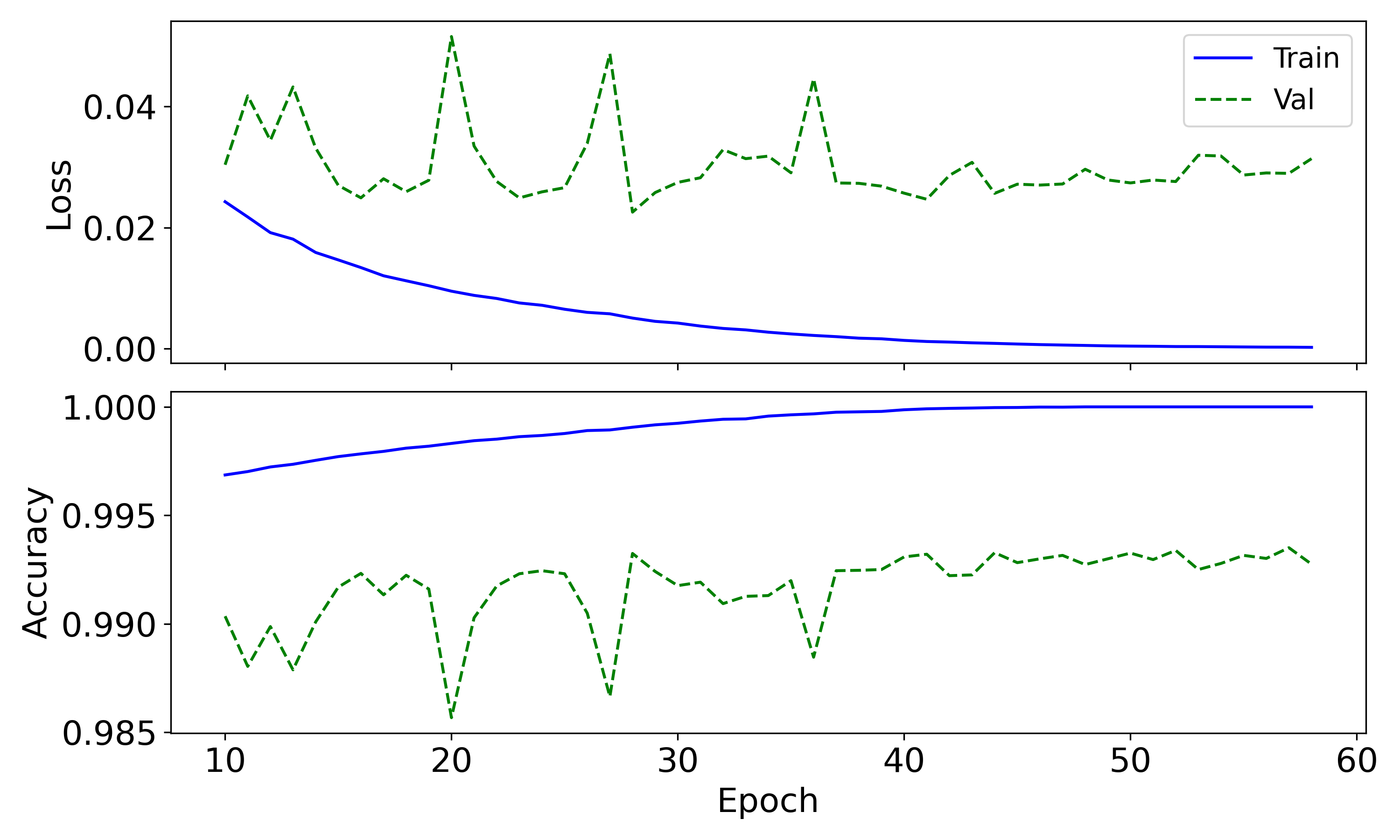}
            \setlength{\height}{\heightof{\includegraphics{Figures/ex_cnn_cls/ex_cls_cnn_5_loss.png}}}
            \put(0.125\columnwidth,0.31\height){(c)}
        \end{overpic}
        \phantomcaption
        \label{fig:ex_cls_loss}
    \end{subfigure}
    \caption{Results for $E_x$ for classification with CNN (compare Fig.~\ref{fig:results_cls_cnn}).}
    \label{fig:ex_results_cls_cnn}
\end{figure}

\begin{figure}[ht]
    \centering
    \begin{subfigure}{\columnwidth}
        \centering
        \begin{tabular}{c||ccc}
            \backslashbox{True}{Pred.}  & 0 & 1 & 2 \\ \hline\hline
            0 & 61808 & 119 & 0  \\
            1 & 271 & 659 & 25 \\
            2 & 6 & 61 & 51
        \end{tabular}
        \phantomcaption
        \label{fig:ex_cls_mlp_matrix}
    \end{subfigure}
    \begin{tikzpicture}[overlay]
        \node at (-2.3,0.25) {(a)};
    \end{tikzpicture}
    \vspace{1em}
    \begin{subfigure}{\columnwidth}
        \centering
        \begin{tabular}{cccc}
            \hline\hline
            Class & Precision & Recall & F1-score \\ \hline
            0 & 0.996 & 0.998 & 0.997 \\
            1 & 0.786 & 0.690 & 0.735 \\
            2 & 0.671 & 0.432 & 0.526 \\
            \hline\hline
        \end{tabular}
        \phantomcaption
        \label{tab:ex_report_mlp}
    \end{subfigure}
    \vspace{1em}
    \begin{tikzpicture}[overlay]
        \node at (-2.3,0.65) {(b)};
    \end{tikzpicture}
    \begin{subfigure}{\columnwidth}
        \begin{overpic}[width=\columnwidth]{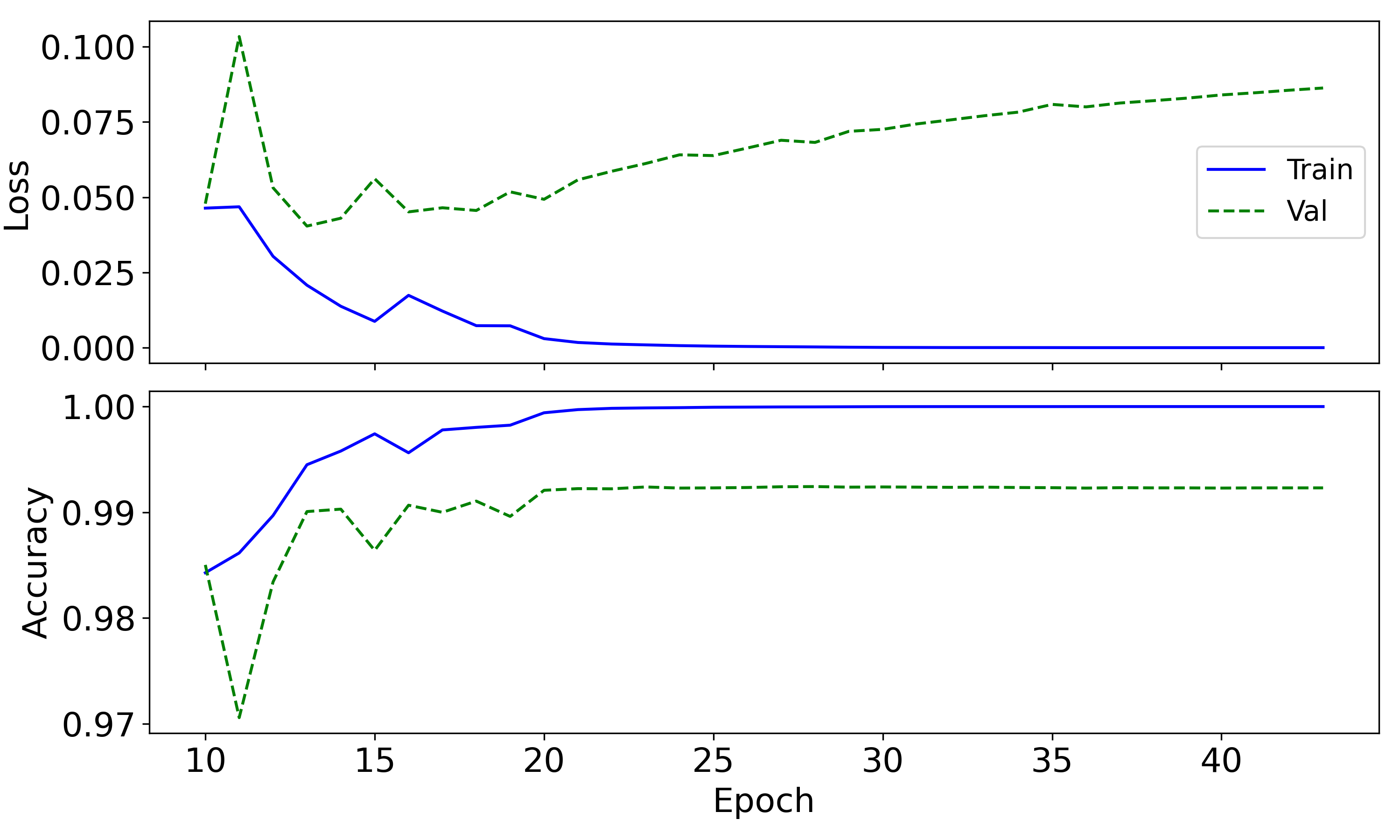}
            \setlength{\height}{\heightof{\includegraphics{Figures/ex_mlp_cls/ex_cls_mlp_5_loss.png}}}
            \put(0.115\columnwidth,0.31\height){(c)}
        \end{overpic}
        \phantomcaption
        \label{fig:ex_cls_mlp_loss}
    \end{subfigure}
    
    \caption{Results for $E_x$ for classification with MLP (compare Fig.~\ref{fig:results_cls_cnn}).}
    \label{fig:ex_results_cls_mlp}
\end{figure}

Figures \ref{fig:ex_results_cls_cnn} and \ref{fig:ex_results_cls_mlp} show results for classification with CNN and MLP models, respectively. They should be compared with the results obtained with the momentum data (see Figures~\ref{fig:results_cls_cnn} and \ref{fig:results_cls_mlp} and discussion in Section~\ref{sec:momentum_class}).
The loss stabilization during training and validation of the CNN takes about 40 epochs, Figure~\ref{fig:ex_cls_loss}, which is twice as fast compared to
the momentum input data. For the MLP network, this process is even quicker, taking half the time, as illustrated in Figure~\ref{fig:ex_cls_mlp_loss}. but the validation loss continuously grows, indicating overfitting. As observed from the performance metrics in the tables in Figure~\ref{fig:ex_results_cls_cnn}a-b and~\ref{tab:ex_report_mlp}a-b, the results for both NN are similar. The classification of class 0 particles performs equally well for both $E_x$ and momentum data, achieving  very high F1-scores with both networks. Class 1 particles are classified with higher precision but lower recall using the electric field data. This results in an F1-score for the CNN that is similar to the one obtained with the momentum data. However, in the case of the MLP, the F1-score is lower due to many class 1 particles being predicted as class 0.
The performance for class 2 particles is sub-optimal, with F1-scores close to 0.5 for both NNs, reflecting equally poor precision and recall metrics. Compared to the results for the momentum data, this is due to about 50\% of class 2 particles being misidentified as class 1, and occasionally as class 0 particles in the case of the MLP.
The lower performance of classification algorithms when operating on electric field data, compared to momentum data, can be attributed to the inherent noise in the electric field measurements and the reliance on a single data dimension. In contrast, the analysis of momentum data benefits from the availability of three independent components, providing richer information and leading to more accurate classification. As a result, classifying high-energy particles based solely one a single component of the electric field is less reliable and it might require other methods or the addition of other input values such as the momentum components.

\subsubsection{Regression}
\begin{figure}[ht]
    \centering
    \begin{subfigure}{\columnwidth}
        \begin{overpic}[width=\columnwidth]{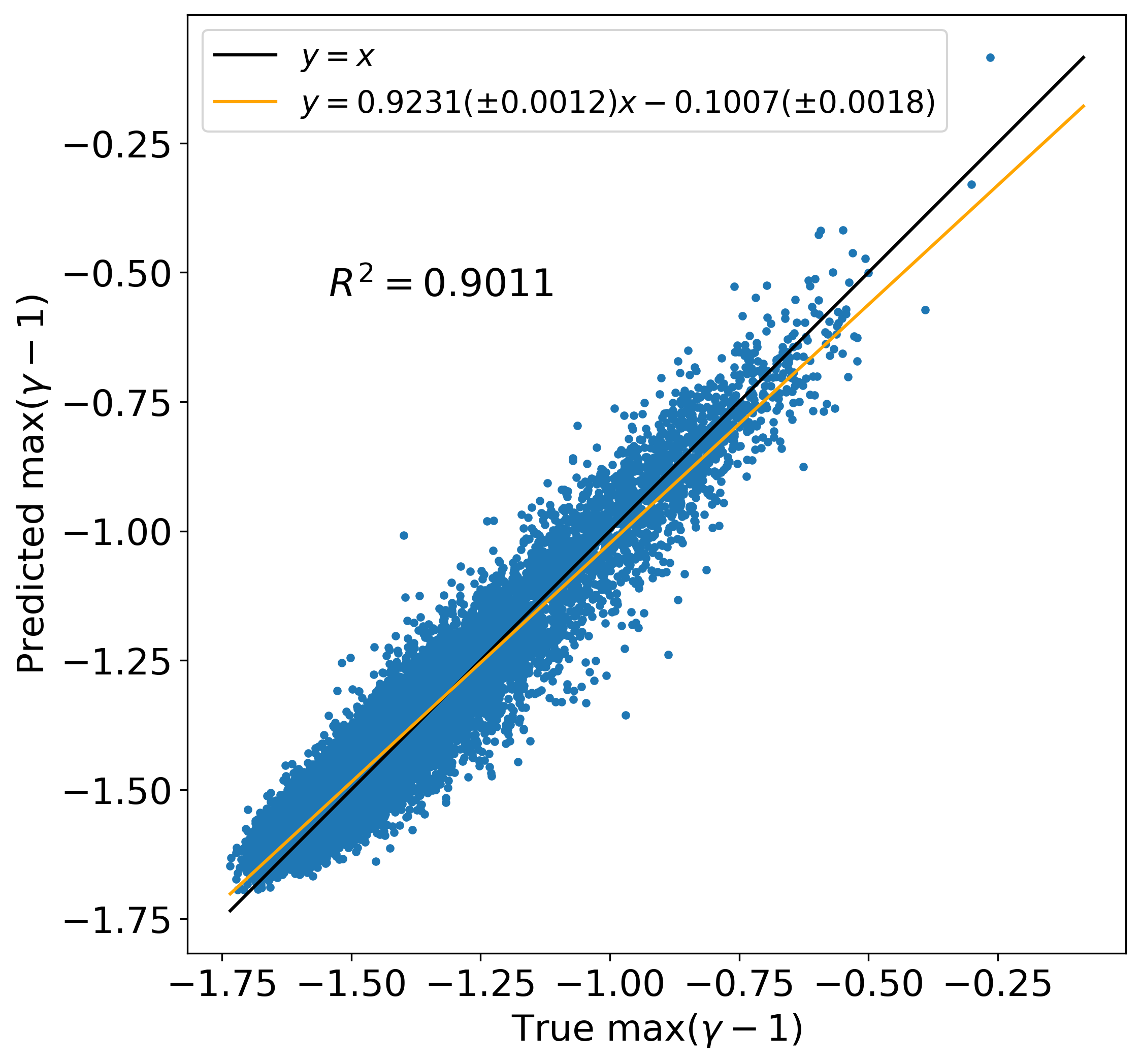}
            \setlength{\height}{\heightof{\includegraphics{Figures/ex_cnn_reg/ex_reg_cnn_5_regression.png}}}
            \put(0.18\columnwidth,0.38\height){(a)}
        \end{overpic}
        \phantomcaption
        \label{fig:ex_plot_cnn}
    \end{subfigure}
    \begin{minipage}[s]{\columnwidth}
        \begin{subfigure}{\columnwidth}
            \begin{overpic}[width=\columnwidth]{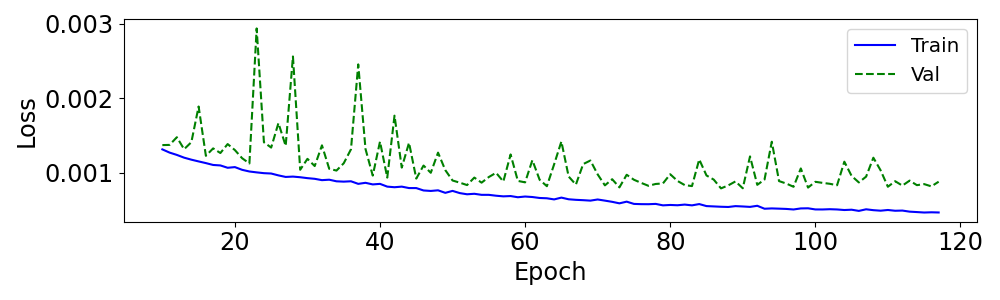}
                \setlength{\height}{\heightof{\includegraphics{Figures/ex_cnn_reg/ex_reg_cnn_5_loss.png}}}
                \put(0.13\columnwidth,0.28\height){(b)}
            \end{overpic}
            \phantomcaption
            \label{fig:ex_loss_cnn_reg}
        \end{subfigure}
        \begin{subfigure}{\columnwidth}
            \begin{overpic}[width=\columnwidth]{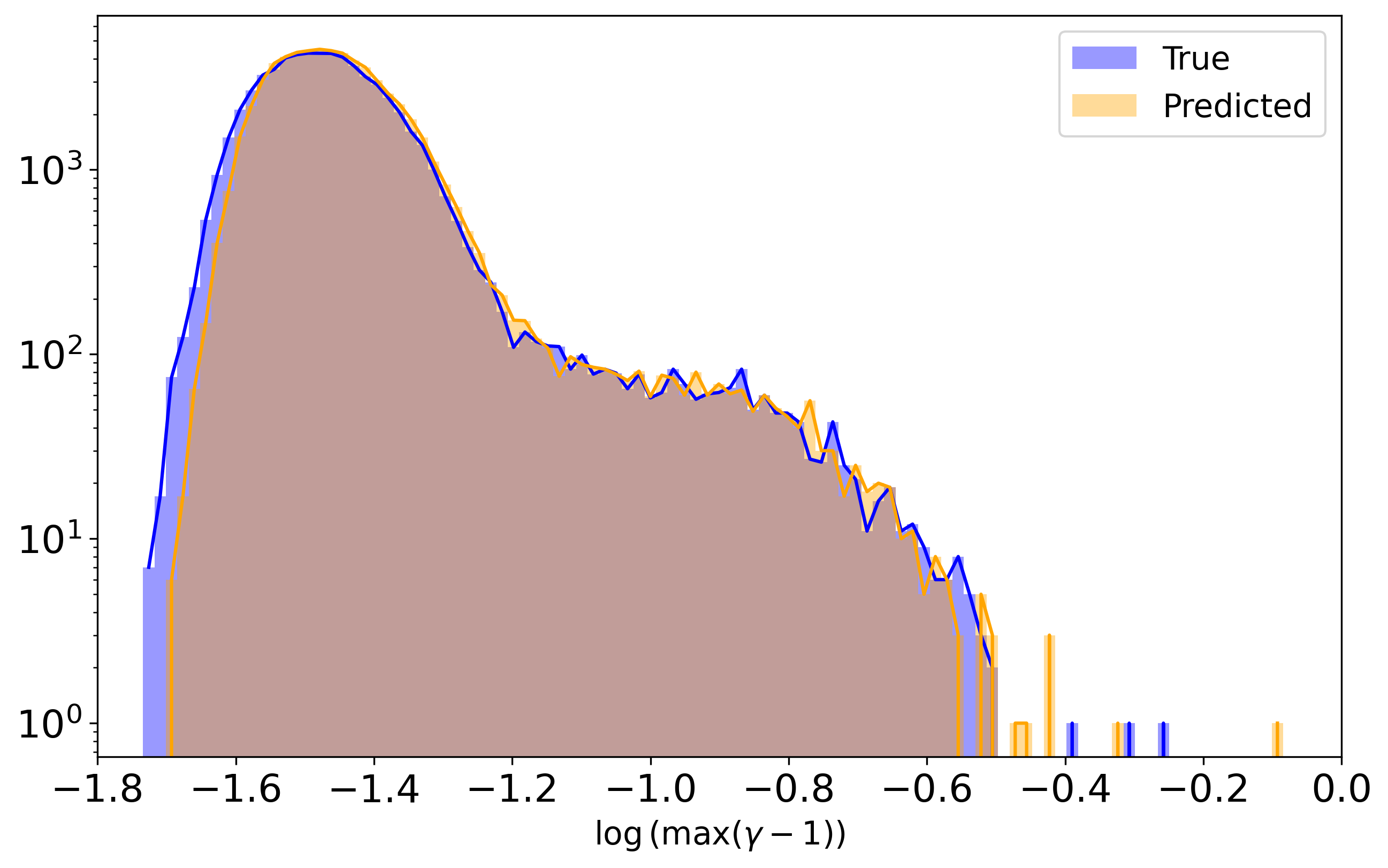}
                \setlength{\height}{\heightof{\includegraphics{Figures/ex_cnn_reg/ex_reg_cnn_5_histogram.png}}}
                \put(0.08\columnwidth,0.36\height){(c)}
            \end{overpic}
            \phantomcaption
            \label{fig:ex_hist_cnn}
        \end{subfigure}
    \end{minipage}
    \caption{Results for regression with CNN (compare Fig.~\ref{fig:results_reg_cnn}).}
    \label{fig:ex_results_reg_cnn}
\end{figure}

\begin{figure}[ht]
    \centering
    \begin{subfigure}{\columnwidth}
        \begin{overpic}[width=\columnwidth]{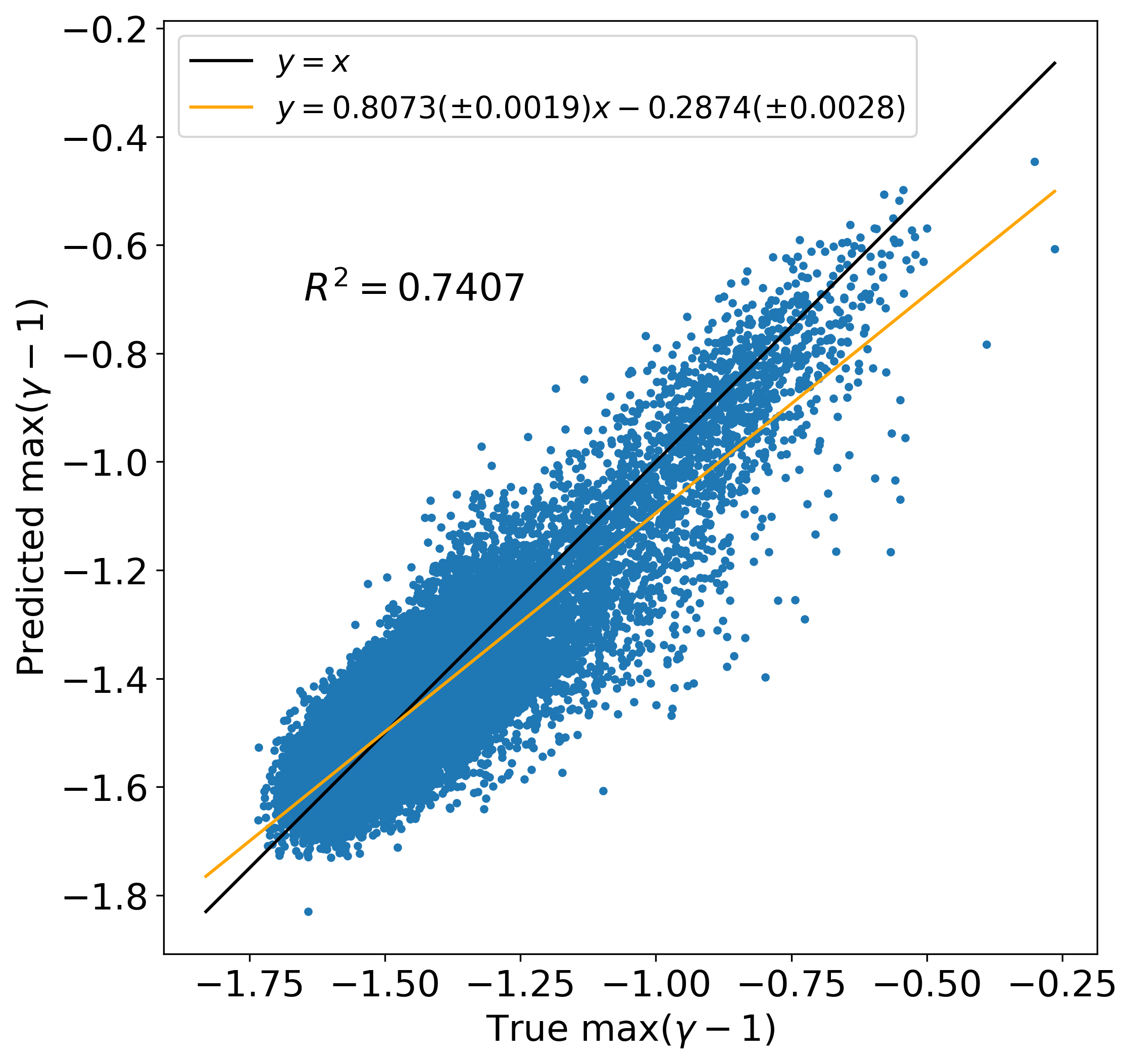}
            \setlength{\height}{\heightof{\includegraphics{Figures/ex_mlp_reg/ex_reg_mlp_4_regression.png}}}
            \put(0.16\columnwidth,0.38\height){(a)}
        \end{overpic}
        \phantomcaption
        \label{fig:ex_plot_mlp}
    \end{subfigure}
    \begin{minipage}[s]{\columnwidth}
        \begin{subfigure}{\columnwidth}
            \begin{overpic}[width=\columnwidth]{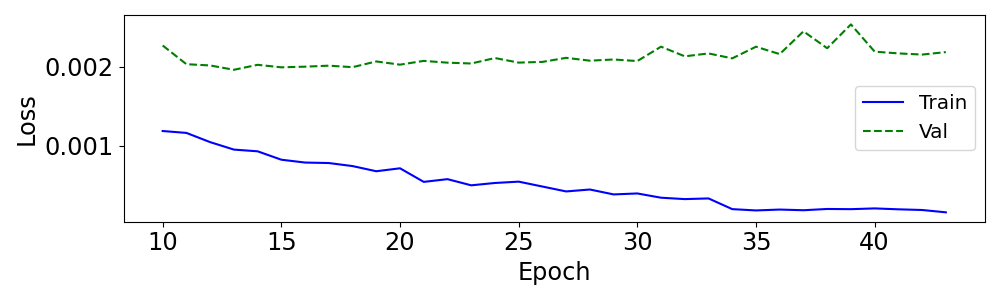}
                \setlength{\height}{\heightof{\includegraphics{Figures/ex_mlp_reg/ex_reg_mlp_4_loss.png}}}
                \put(0.13\columnwidth,0.28\height){(b)}
            \end{overpic}
            \phantomcaption
            \label{fig:ex_loss_mlp_reg}
        \end{subfigure}
        \begin{subfigure}{\columnwidth}
            \begin{overpic}[width=\columnwidth]{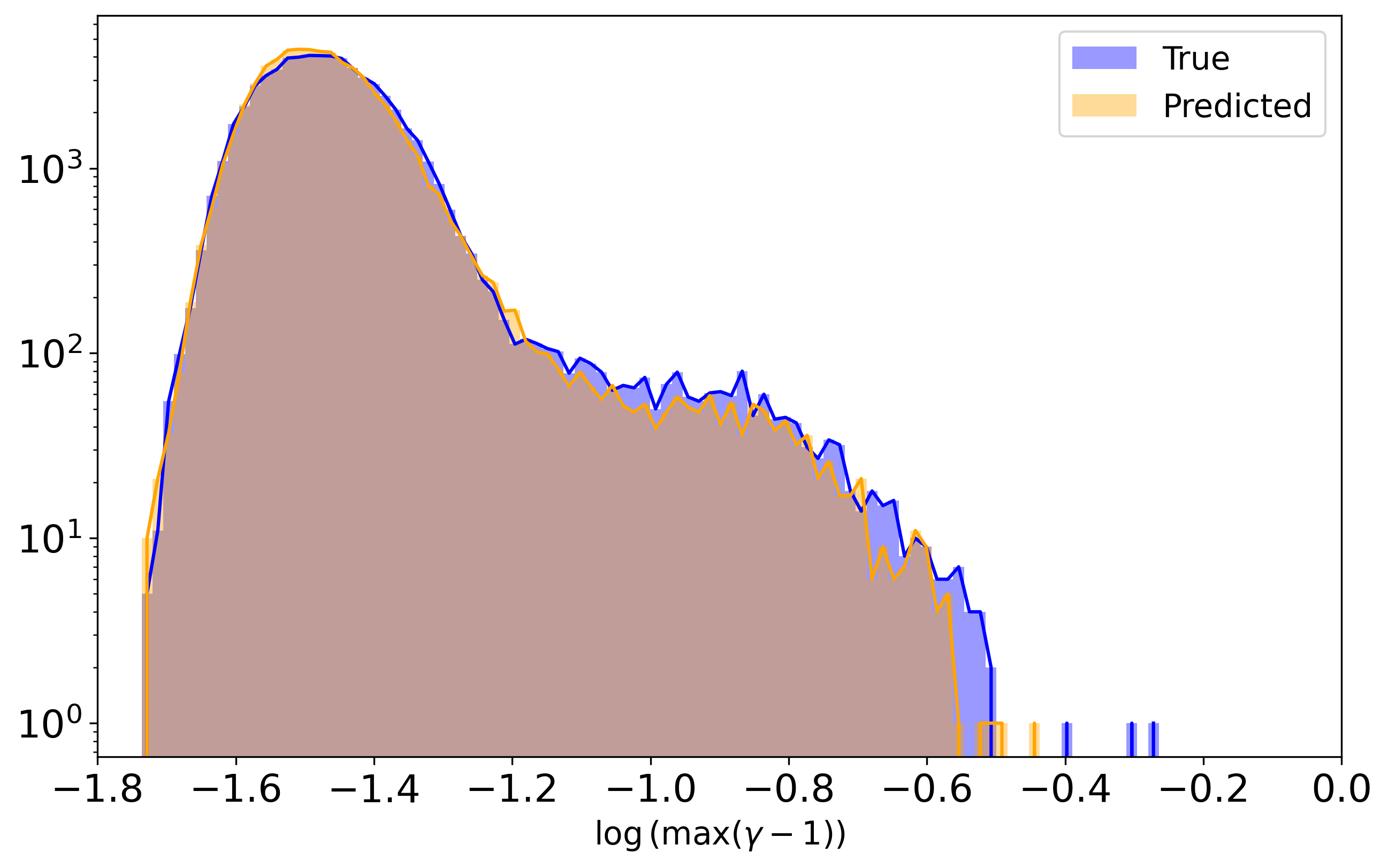}
                \setlength{\height}{\heightof{\includegraphics{Figures/ex_mlp_reg/ex_reg_mlp_4_histogram.png}}}
                \put(0.08\columnwidth,0.36\height){(c)}
            \end{overpic}
            \phantomcaption
            \label{fig:ex_hist_mlp}
        \end{subfigure}
    \end{minipage}
    
    \caption{Results for regression with MLP (compare Fig.~\ref{fig:results_reg_cnn}).}
    \label{fig:ex_results_reg_mlp}
\end{figure}

Figures \ref{fig:ex_results_reg_cnn} and \ref{fig:ex_results_reg_mlp} show results for regression for the CNN and MLP, respectively. As in the case of classification, the CNN performs better than the MLP. 
This is because a
CNN is able to filter noise more effectively by performing convolutions, which smooth the input and reduce noise in the deeper layers of the NN. In both cases, the linear regression is sloped towards lower values due to the very large amount of thermal particles which skew the regression, and exhibits a higher spread, reflected in the lower $R^2$ scores obtained. For the CNN, an $R^2=0.9011$ is achieved, which is notable considering only $E_x$ is used as input data. In contrast, the MLP yields an $R^2=0.7407$, performing significantly worse than the CNN.
Note, that although the histograms in Figures~\ref{fig:ex_hist_cnn} and~\ref{fig:ex_hist_mlp} appear similar, the results, as evidenced by the linear regressions, are less accurate due to the spread of values not being fully captured by the histograms.

\subsubsection{Anomaly Detection}
\begin{figure}[ht]
    \centering
    \begin{subfigure}{\columnwidth}
        \begin{overpic}[width=\columnwidth]{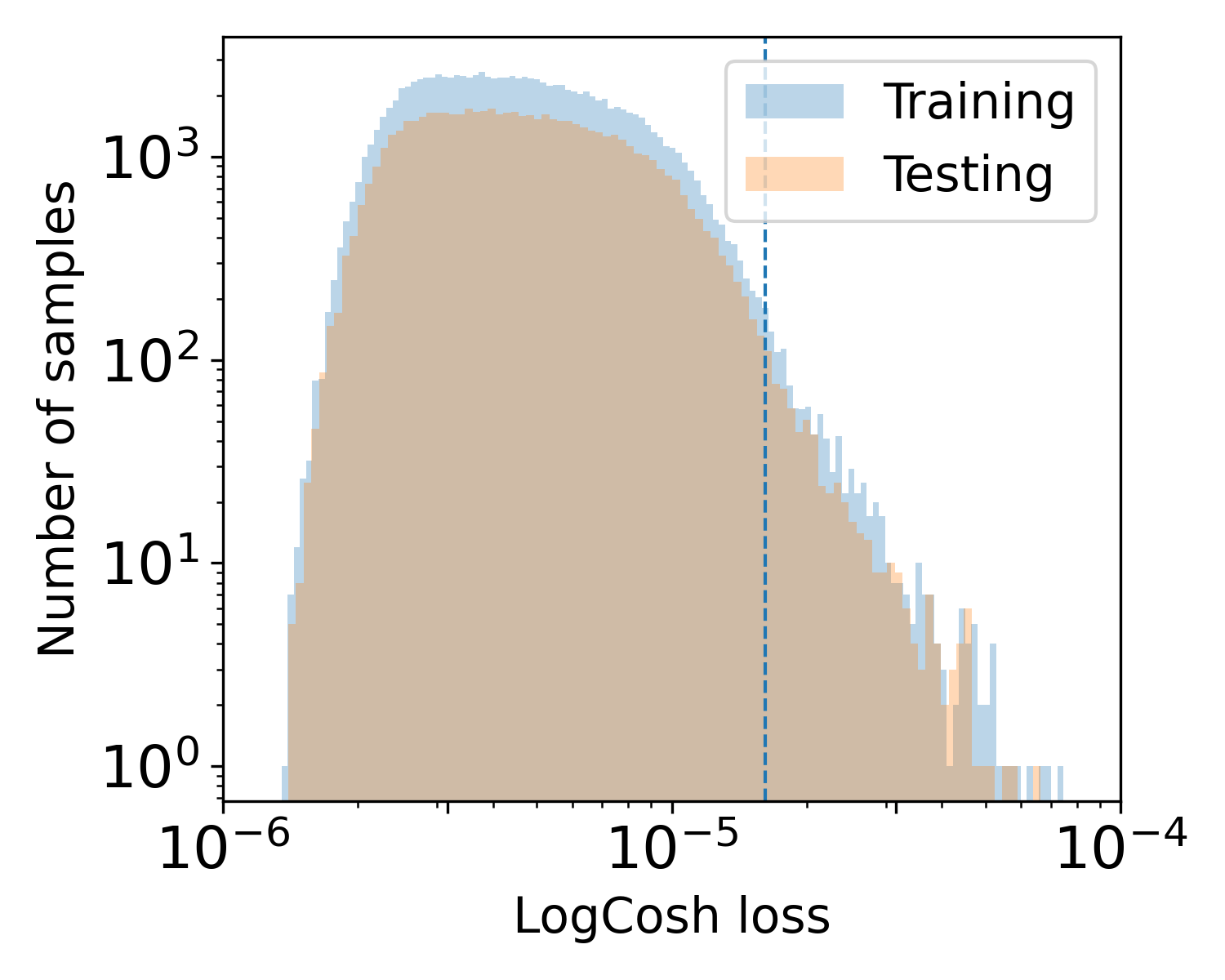}
            \setlength{\height}{\heightof{\includegraphics{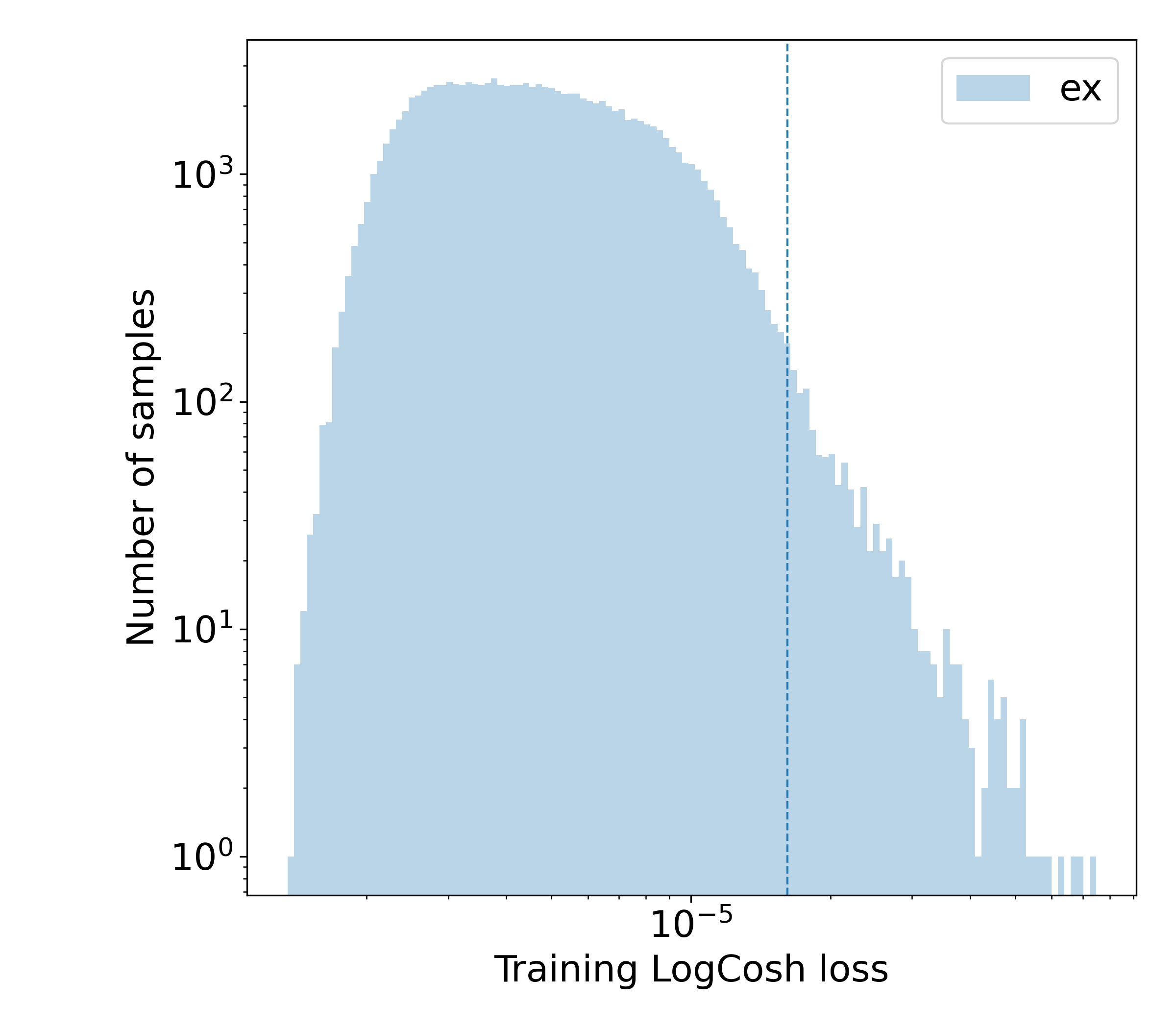}}}
            \put(0.2\columnwidth,0.35\height){(a)}
        \end{overpic}
        \phantomcaption
        \label{fig:loss_ex}
    \end{subfigure}
    \begin{subfigure}{\columnwidth}
        \begin{overpic}[width=\columnwidth]{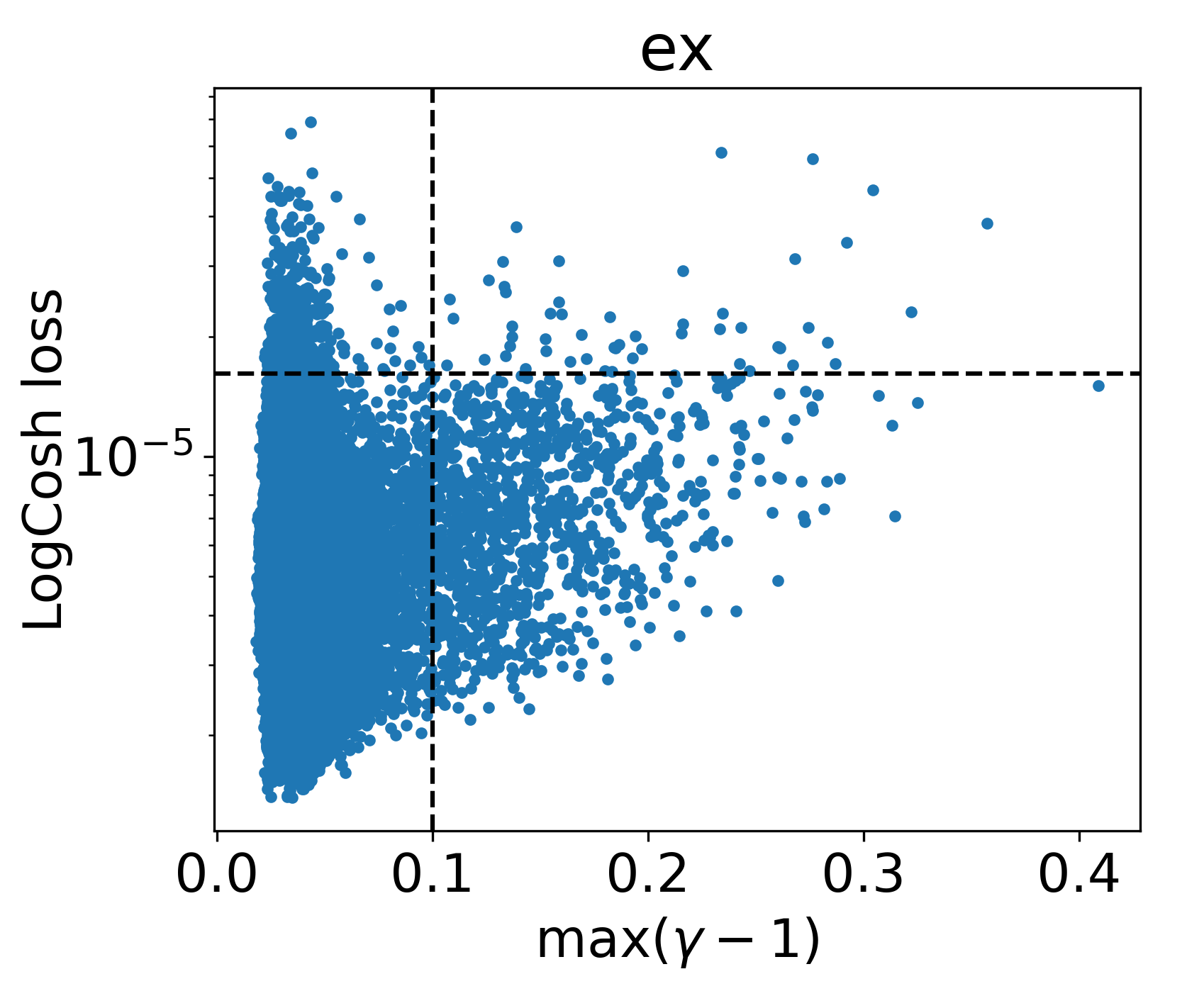}
            \setlength{\height}{\heightof{\includegraphics{Figures/ad_ex/ex_ad_bottleneck_007_008_1_anomaly_detection_label_vs_loss.png}}}
            \put(0.19\columnwidth,0.52\height){(b)}
        \end{overpic}
        \phantomcaption
        \label{fig:label_vs_loss_ex}
    \end{subfigure}
    \begin{subfigure}{\columnwidth}
            \centering
            \begin{tabular}{c|c c}
                 & \begin{tabular}{c} Predicted\\Non-anomaly\end{tabular} & \begin{tabular}{c} Predicted\\Anomaly\end{tabular}\\ \hline
              \begin{tabular}{c}Actual\\Non-anomaly\end{tabular} & 60917 & 592 \\
             \begin{tabular}{c}Actual\\Anomaly\end{tabular} & 1429 & 62 \\
            \end{tabular}
            \phantomcaption
            \label{tab:ad_matrix_ex}
            \begin{tikzpicture}[overlay]
                \node at (-5.,0.9) {(c)};
            \end{tikzpicture}
    \end{subfigure}
    \caption{Results for anomaly detection with CNN for the $E_x$ electric field (compare Fig.~\ref{fig:results_reg_cnn}).}
    \label{fig:results_ad_ex}
\end{figure}

Similar to the approach taken with momentum data for Anomaly Detection, we trained the model using only low-energy particles for the electric field data. Results of this analysis are presented in Figure~\ref{fig:results_ad_ex}. An additional analysis using the full dataset for training was conducted; however, it delivered significantly worse performance compared to training with only low-energy particles and is therefore not reported here.

As in the case of the momentum data with the full dataset, for AD with the $E_x$ electric field we set the threshold at 1\% of the full dataset. The training and testing loss histograms in Figure~\ref{fig:loss_ex} are very similar, which indicates that the NN struggles to distinguish between non-energetic and energetic particles effectively. This is due to the electric field time series being noisier compared to momentum data, making it difficult for the network to discern the non-thermal pattern. As a result, the performance is worse compared to when momentum data is used. 
Figure \ref{fig:label_vs_loss_ex} clearly illustrates that even particles with low energies can exhibit high loss due to a wide range of field behaviours among thermal particles. At the same time, most of energetic particles are misclassified as non-anomalies. This is quantified in the confusion matrix in Figure~\ref{tab:ad_matrix_ex}, which yields a very poor F1-score of 5.8\%. Clearly then, the AD employed with the electric field data is not an adequate tool to analyse PIC simulation particle time series.

\section{Conclusions}
We employ neural networks to predict the maximum kinetic energy of particles passing through the Buneman instability region, where some of these particles undergo SSA. Using all three components of momentum as input data, supervised learning delivers impressive results despite the highly non-linear nature of the dataset. In the case of classification, both convolutional neural network (CNN) and multi-layer perceptron (MLP) demonstrate comparable performance, yet CNN outperforms in terms of precision, recall, and most notably, F1-score. In the case of regression, both CNN and MLP perform very similarly, as indicated by the goodness-of-fit and thus both are suitable for regression analysis. Finally, anomaly detection successfully identifies energetic particles without requiring a class or label, as is the case of supervised learning. We employed a semi-anomaly detection approach, using only the thermal population as the training dataset, as well as a full anomaly detection approach, utilising the entire dataset. The results are notably worse than those from classification and regression, primarily due to the absence of labels, which makes prediction more challenging for the neural network. In the case of the full training dataset approach, the challenge is even greater, as the network must contend with the entire dataset.

Alternatively, we also used the electric field in $x-$direction, $E_x$, as input data and applied the same algorithms and types of neural networks. It is noteworthy that the electric field, as shown in Figure~\ref{fig:bun_traj}, is a highly noisy dataset. Consequently, we expect the neural network to face challenges in correctly predicting the values of the maximum kinetic energy, resulting in comparatively worse outcomes than when using momentum data. Surprisingly, despite the inherent noise in the electric field, the results are quite acceptable. In the context of classification, both CNN and MLP achieved reasonable results, as evidenced by the relatively diagonal confusion matrices. In regression analysis, the performance was particularly impressive for CNN, as indicated by the goodness-of-fit. However, unlike in classification, MLP performs worse than CNN, and linear regression showed clearly inferior results. Therefore, we consider CNN as the most suitable neural network architecture for handling noisy data in future applications. However, anomaly detection with $E_x$ yielded results significantly worse than those obtained with momentum. The neural network is unable to reliably distinguish between energetic and non-energetic particles. The time series for $E_x$ were too noisy, and we therefore discourage the use of AD for such noisy time series.

A PIC simulation can be paused at any time step, enabling particle tracing to begin in subsequent steps. Investigating the microphysics of acceleration processes often requires tracking a large number of particles, which can be computationally expensive. When particle energization occurs in stages involving different, interdependent acceleration mechanisms, neural networks offer an efficient method for identifying energetic particles likely to undergo further energization. By selectively tracing these particles, computational costs are significantly reduced, making this approach particularly advantageous for large-scale simulations.
Plasma shock systems where the Buneman instability is excited provide an example of such multi-stage particle energization. Particles pre-accelerated via SSA in the Buneman instability region are often subjected to further acceleration in the shock ramp and downstream, through processes such as drift shock acceleration (DSA \cite{DSA}) or second-order Fermi acceleration\cite{BohdanSSA}.

Our results show that, in the scenario of particles interacting with Buneman instability waves, we are able to predict their energies using only particle momentum or even a single electric field component time series. The different methods we tested yield varying outcomes depending on the chosen input variables, providing valuable insights for further development and the investigation of other particle acceleration scenarios. As computer simulations continue to grow in size, neural network-based tools are expected to facilitate faster and more precise analysis of simulation data, particularly for traced particles in fully kinetic PIC or hybrid kinetic simulations.

\begin{acknowledgments}
    This work has been supported by Narodowe Centrum Nauki through research project 2019/33/B/ST9/02569. A.B. was supported by the German Research Foundation (DFG) as part of the Excellence Strategy of the federal and state governments - EXC 2094 - 390783311. We gratefully acknowledge Polish high-performance computing infrastructure PLGrid (HPC Center: ACK Cyfronet AGH) for providing computer facilities and support within computational grant no. PLG/2023/016378 and also the North German Supercomputing Alliance (HLRN) under the project bbp00033.
\end{acknowledgments}

\section*{AUTHOR DECLARATIONS}

\subsection*{Conflict of Interest}
The authors have no conflicts to disclose.

\subsection*{Author Contributions}
\textbf{Gabriel Torralba Paz:} Conceptualization (supporting); Formal Analysis (lead); Methodology (lead); Software (lead); Visualization (lead); Writing --- original draft (lead); Writing --- Review \& editing (supporting). \textbf{Artem Bohdan:} Conceptualization (lead); Methodology (supporting); Software (supporting); Validation (lead); Visualization (supporting); Writing --- Review \& editing (supporting). \textbf{Jacek Niemiec:} Supervision (lead); Resources (lead); Visualization (supporting); Writing --- Review \& editing (lead)

\section*{DATA AVAILABILITY}
The data that support the findings of this study are available from the corresponding author upon reasonable request.

\appendix*

\section{Neural Networks} \label{sec:appendix}
Here we present the neural networks used for the analysis of our dataset. 
Each network is displayed in a table format, listing the layers and their specifications. The network structure starts from the top (Input) and progresses to the bottom (Output) of the table. A plus sign denotes a sequence of layers that are applied consecutively, as explained in Section~\ref{sec:cnn}.
A convolutional layer is represented as "(output size, filters, kernel size)", where "output size" indicates the size of the array after the input data passes through the convolutional layer, "filters" refers to the number of filters/units in the layer, and "kernel size" is the size of the convolutional window that performs a convolutional operation on the input array. The leak parameter of the Leaky ReLU is denoted as $\alpha$. A dense layer (a deeply connected layer used in MLP) is defined by a single value, "units", which is analogous to filters in a convolutional layer.

\begin{table*}[hbt!]
    \centering
    \begin{tabular}{c|c|c|c}
        \hline\hline
        Layers & \multicolumn{3}{c}{Specifications} \\
        \hline
        Input & $p_x(1200, 1)$ & $p_y(1200, 1)$ & $p_z(1200, 1)$\\
        Conv+Norm+L.ReLU 1 & $(600, 12, 5), \alpha=0.0268$ & $(300, 13, 5), \alpha=0.0244$ & $(400, 30, 5), \alpha=0.017$ \\
        Conv+Norm+L.ReLU 2 & $(150, 26, 4), \alpha=0.0158$ & $(150, 10, 5), \alpha=0.0247$ & $(133, 29, 4), \alpha=0.033$ \\
        Conv+Norm+L.ReLU 3 & $(150, 20, 5), \alpha=0.0269$ & $( 50, 20, 3), \alpha=0.0226$ & $(133, 21, 4), \alpha=0.033$ \\
    Max Pooling & 20 & 20 & 21 \\
        Flatten & 20 & 20 & 21 \\
    Concatenate & \multicolumn{3}{c}{61} \\
        Output & \multicolumn{3}{c}{Softmax(3), LR = 0.000575} \\
        \hline\hline
    \end{tabular}
    \caption{Classification algorithm with a convolutional neural network (CNN). The layers are represented as (output size, filters, kernel size), and $\alpha$ is the leak parameter of the Leaky ReLU. LR is the learning rate of the neural network.} 
    \label{tab:cls_cnn}
\end{table*}

\begin{table*}[hbt!]
    \centering
    \begin{tabular}{c|c|c|c}
        \hline\hline
        Layers & \multicolumn{3}{c}{Specifications} \\
        \hline
        Input & $p_x(1200)$ & $p_y(1200)$ & $p_z(1200)$ \\
        Dense+L.ReLU & $(17, \alpha=0.0292)$ & $(163, \alpha=0.0149)$ & $(47, \alpha=0.0297)$ \\
        Dense+L.ReLU & $(187, \alpha=0.0234)$ & $(126, \alpha=0.0302)$ & $(221, \alpha=0.0365)$ \\
        Dense+L.ReLU & $(89, \alpha=0.0372)$ & $(72, \alpha=0.0395)$ & $(53, \alpha=0.0203)$ \\
        Concatenate & \multicolumn{3}{c}{214} \\
        Output & \multicolumn{3}{c}{Softmax(3), LR = 0.00037} \\
        \hline\hline
    \end{tabular}
    \caption{Classification algorithm with a multi-layer perceptron (MLP). The layers are represented as (units) and $\alpha$ is the leak parameter of the Leaky ReLU. LR is the learning rate of the neural network.}
    \label{tab:cls_mlp}
\end{table*}

\begin{table*}[hbt!]
    \centering
    \begin{tabular}{c|c|c|c}
        \hline\hline
        Layers & \multicolumn{3}{c}{Specifications} \\
        \hline
        Input & $p_x(1200, 1)$ & $p_y(1200, 1)$ & $p_z(1200, 1)$ \\
        Conv+Norm+L.ReLU 1 & $(400, 16, 4), \alpha=0.03$ & $(400, 16, 4), \alpha=0.03$ & $(400, 16, 4), \alpha=0.03$ \\
        Conv+Norm+L.ReLU 2 & $(200, 16, 4), \alpha=0.03$ & $(200, 16, 4), \alpha=0.03$ & $(200, 16, 4), \alpha=0.03$ \\
        Conv+Norm+L.ReLU 3 & $(100, 32, 4), \alpha=0.03$ & $(100, 32, 4), \alpha=0.03$ & $(100, 32, 4), \alpha=0.03$ \\
        Conv+Norm+L.ReLU 4 & $(50, 64, 4), \alpha=0.03$ & $(50, 64, 4), \alpha=0.03$ & $(50, 64, 4), \alpha=0.03$ \\
        Conv+Norm+L.ReLU 5 & $(25, 64, 4), \alpha=0.03$ & $(25, 64, 4), \alpha=0.03$ & $(25, 64, 4), \alpha=0.03$ \\
    Max Pooling & 64 & 64 & 64 \\
        Flatten & 64 & 64 & 64 \\
    Concatenate & \multicolumn{3}{c}{192} \\
        Output & \multicolumn{3}{c}{Linear(1)} \\
        \hline\hline
    \end{tabular}
    \caption{Regression algorithm with a CNN (see Table~\ref{tab:cls_cnn}).
    }
    \label{tab:reg_cnn}
\end{table*}

\begin{table*}[hbt!]
    \centering
    \begin{tabular}{c|c|c|c}
        \hline\hline
        Layers & \multicolumn{3}{c}{Specifications} \\
        \hline
        Input & $p_x(1200)$ & $p_y(1200)$ & $p_z(1200)$ \\
        Dense+L.ReLU & $(302, \alpha=0.0309)$ & $(37, \alpha=0.0181)$ & $(99, \alpha=0.0332)$ \\
        Dense+L.ReLU & $(63, \alpha=0.024)$ & $(206, \alpha=0.0368)$ & $(69, \alpha=0.0147)$ \\
        Dense+L.ReLU & $(372, \alpha=0.0278)$ & $(166, \alpha=0.0296)$ & $(271, \alpha=0.0186)$ \\
        Dense+L.ReLU & $(486, \alpha=0.0248)$ & $(285, \alpha=0.0185)$ & $(259, \alpha=0.0133)$ \\
    Concatenate & \multicolumn{3}{c}{1030} \\
            Output & \multicolumn{3}{c}{Linear(1), LR = 0.000439} \\
            \hline\hline
    \end{tabular}
    \caption{Regression algorithm with a MLP (see Table~\ref{tab:cls_mlp}). 
    }
    \label{tab:reg_mlp}
\end{table*}

\begin{table}[hbt!]
    \centering
    \begin{tabular}{c|c}
        \hline\hline
        \multicolumn{1}{c|}{Layers} & Specifications \\
        \hline
        \multicolumn{1}{c|}{Input} & $(1200[p_x, p_y, p_z], 3)$\\
        Conv+Norm+L.ReLU($0.03$) & (1200, 64, 3) \\
        Conv+Norm+L.ReLU($0.03$) & (1200, 32, 3) \\
        Conv+Norm+L.ReLU($0.03$) & (1200, 16, 3) \\
        Conv+Norm+L.ReLU($0.03$) & (1200, 8 , 3) \\      
        Conv+Norm+L.ReLU($0.03$) & (1200, 4 , 3) \\
        Deconv+Norm+L.ReLU($0.03$) & (1200, 4 , 3) \\
        Deconv+Norm+L.ReLU($0.03$) & (1200, 8 , 3) \\
        Deconv+Norm+L.ReLU($0.03$) & (1200, 16, 3) \\
        Deconv+Norm+L.ReLU($0.03$) & (1200, 32, 3) \\      
        Deconv+Norm+L.ReLU($0.03$) & (1200, 64, 4) \\
        Deconv (Output) & (1200, 3), LR=0.0005 \\
        \hline\hline
    \end{tabular}
    \caption{Anomaly detection with a CNN. The layers are convolutional neural network as (output size, filters, kernel size).}
    \label{tab:ad_cnn}
\end{table}

\begin{table}[hbt!]
    \centering
    \begin{tabular}{c|c}
        \hline\hline
        Layers & Specifications \\
        \hline
        Input & $E_x(1200, 1)$\\
        Conv+Norm+L.ReLU 1 & $(400, 16, 5), \alpha=0.03$\\
        Conv+Norm+L.ReLU 2 & $(200, 16, 5), \alpha=0.03$\\
        Conv+Norm+L.ReLU 3 & $(100, 32, 5), \alpha=0.03$\\
        Conv+Norm+L.ReLU 3 & $(50, 64, 5), \alpha=0.03$\\
        Conv+Norm+L.ReLU 3 & $(25, 64, 5), \alpha=0.03$\\
        Max Pooling & 64 \\
        Output & Softmax(3), LR = 0.000932 \\
        \hline\hline
    \end{tabular}
    \caption{Classification algorithm with a CNN (see Table~\ref{tab:cls_cnn}). 
    }
    \label{tab:cls_cnn_ex}
\end{table}

\begin{table}[hbt!]
    \centering
    \begin{tabular}{c|c}
        \hline\hline
        Layers & Specifications \\
        \hline
        Input & $E_x(1200, 1)$\\
        Dense+L.ReLU & $(218, \alpha=0.0154)$\\
        Dense+L.ReLU & $(247, \alpha=0.0299)$\\
        Dense+L.ReLU & $(226, \alpha=0.0327)$\\
        Dense+L.ReLU & $(247, \alpha=0.0181)$\\
        Dense+L.ReLU & $(114, \alpha=0.0287)$\\
        Output & Softmax(3), LR = 0.000933 \\
        \hline\hline
    \end{tabular}
    \caption{Classification algorithm with a MLP (see Table~\ref{tab:cls_mlp}).
    }
    \label{tab:cls_mlp_ex}
\end{table}

\begin{table}[hbt!]
    \centering
    \begin{tabular}{c|c}
        \hline\hline
        Layers & Specifications \\
        \hline
        Input & $E_x(1200, 1)$\\
        Conv+Norm+L.ReLU 1 & $(300, 72, 4), \alpha=0.0101$\\
        Conv+Norm+L.ReLU 2 & $(150, 40, 6), \alpha=0.0129$\\
        Conv+Norm+L.ReLU 3 & $(30, 28, 5), \alpha=0.0153$\\
        Conv+Norm+L.ReLU 3 & $(10, 53, 7), \alpha=0.0311$\\
        Conv+Norm+L.ReLU 3 & $(5, 38, 7), \alpha=0.0386$\\
        Output & Linear(1), LR = 0.000505 \\
        \hline\hline
    \end{tabular}
    \caption{Classification algorithm with a CNN (see Table~\ref{tab:cls_cnn}). 
    }
    \label{tab:reg_cnn_ex}
\end{table}

\begin{table}[hbt!]
    \centering
    \begin{tabular}{c|c}
        \hline\hline
        Layers & Specifications \\
        \hline
        Input & $E_x(1200, 1)$\\
        Dense+L.ReLU & $(499, \alpha=0.0273)$\\
        Dense+L.ReLU & $(499, \alpha=0.0172)$\\
        Dense+L.ReLU & $(298, \alpha=0.0112)$\\
        Dense+L.ReLU & $(386, \alpha=0.0102)$\\
        Output & Linear(1), LR = 0.000861 \\
        \hline\hline
    \end{tabular}
    \caption{Classification algorithm with a MLP (see Table~\ref{tab:cls_mlp}).
    }
    \label{tab:reg_mlp_ex}
\end{table}

\begin{table}[hbt!]
    \centering
    \begin{tabular}{c|c}
        \hline\hline
        \multicolumn{1}{c|}{Layers} & Specifications \\
        \hline
        \multicolumn{1}{c|}{Input} & $(1200[E_x], 1)$\\
        Conv+Norm+L.ReLU($0.03$) & (1200, 64, 3) \\
        Conv+Norm+L.ReLU($0.03$) & (1200, 32, 3) \\
        Conv+Norm+L.ReLU($0.03$) & (1200, 16, 3) \\
        Conv+Norm+L.ReLU($0.03$) & (1200, 8, 3) \\      
        Deconv+Norm+L.ReLU($0.03$) & (1200, 16, 3) \\
        Deconv+Norm+L.ReLU($0.03$) & (1200, 32, 3) \\
        Deconv+Norm+L.ReLU($0.03$) & (1200, 64, 3) \\
        Deconv (Output) & (1200, 1), LR=0.000575 \\
        \hline\hline
    \end{tabular}
    \caption{Anomaly detection with a CNN (see Table~\ref{tab:ad_cnn}).
    }
    \label{tab:ad_cnn_ex}
\end{table}

\bibliography{sample}

\FloatBarrier

\end{document}